\documentclass[10pt]{article}

\usepackage{graphicx}
\usepackage{dcolumn}
\usepackage{bm}
\usepackage{amsmath}
\usepackage[caption=false,font=footnotesize]{subfig}
\usepackage{cite}

\newcommand{\V}{\bm}
\newcommand{\photonsperpulsemm}{\mathrm{photons}/\mathrm{pulse}/\mathrm{mm}^2}

\advance \oddsidemargin by -\textwidth
\divide  \oddsidemargin by 2
\advance \oddsidemargin by -1in
\evensidemargin = \oddsidemargin

\newlength{\figurewidth}
\figurewidth = \textwidth

\title{Phase retrieval from single biomolecule diffraction pattern}

\author{Shiro Ikeda\footnote{The Institute of Statistical Mathematics,
    Tokyo, Japan, \tt{shiro@ism.ac.jp}} \and
  Hidetoshi Kono\footnote{Japan Atomic Energy Agency, Kyoto, Japan,
    \tt{kono.hidetoshi@jaea.go.jp}}}

\begin{document}
\maketitle

\begin{abstract}
  In this paper, we propose the SPR (sparse phase retrieval) method,
  which is a new phase retrieval method for coherent x-ray diffraction
  imaging (CXDI). Conventional phase retrieval methods effectively
  solve the problem for high signal-to-noise ratio measurements, but
  would not be sufficient for single biomolecular imaging which is
  expected to be realized with femto-second x-ray free electron laser
  pulses. The SPR method is based on the Bayesian statistics. It does
  not need to set the object boundary constraint that is required by
  the commonly used hybrid input-output (HIO) method, instead a prior
  distribution is defined with an exponential distribution and used
  for the estimation. Simulation results demonstrate that the proposed
  method reconstructs the electron density under a noisy condition
  even some central pixels are masked.
\end{abstract}

\section{Introduction}
\label{sec:introduction}

X-ray free electron lasers (XFELs) can potentially provide us a novel
mean to determine the three-dimensional (3D) structure of biomolecules
from the diffraction images of single
molecules~\cite{Sayre1980,GaffneyChapman2007science}. Conventionally,
x-ray crystallography has been the principal tool to determine
high-resolution 3D structures of proteins, nucleic acids, and their
complexes. However, a critical process is the crystallization, where a
sufficiently large crystal must be prepared. It is known that about
40\% of biomolecules, particularly membrane proteins, do not
crystallize. The coherent x-ray diffraction imaging (CXDI) by XFELs
does not require a crystal and will change the situation.

In order to realize the single molecule imaging, a straightforward
scenario has been
proposed~\cite{Neutze_etal2000nature,HuldtSzokeHajdu2003} and
demonstrated with single mimivirus
particles~\cite{Seibert_etal2011nature}. Firstly, a series of single
molecules with the same structure are injected into vacuum and exposed
to a very short (less than 5~$\mathrm{fs}$) pulse x-ray beam. The
diffraction image of the molecule with unknown orientation is recorded
before radiation-induced structural degradation. This process is
repeated until a sufficient number of images are recorded. The
diffraction data are then classified into groups according to the
orientations of the molecule and averaged to improve the
signal-to-noise
ratio~\cite{BortelFaigel2007jsb,BortelFaigelTegze2009jsb}. Next, the
relative orientations of the classes are determined to assemble a 3D
coherent diffraction pattern in reciprocal space. Finally, the 3D
electron density of the molecule is obtained by a phase retrieval
method.

The above scenario was proved to be effective for the 3D
reconstruction of mimivirus
particles~\cite{Seibert_etal2011nature}. However, the condition for a
single molecule CXDI will be more severe since the number of the
photons received by each pixel will be very limited. Some methods of
reconstructing 3D diffraction intensity have been proposed for this
problem~\cite{Fung_etal2009natphys,LohVeit2009pre,Saldin_etal2009iop}. They
are likely to be computationally intensive and the feasibility remains
to be seen.

Another scenario might be possible. After collecting a sufficient
number of diffraction images with unknown orientations, 2D electron
density images are first reconstructed from the 2D diffraction data,
followed by the 3D reconstruction method used in transmission electron
microscopy~\cite{Frank2006oup}. In either scenario, the difficulty is
due to the small number of photons scattered by a single molecule even
though x-ray beams that linac coherent light source (LCLS) has started
to commission give us illumination which is around $10^9$ times
brighter (${\sim}10^{19}$ $\photonsperpulsemm$ in the unfocused beam)
than that of current synchrotrons~\cite{Young_etal2010nature}.

We propose a new method for retrieving phases of a diffraction image
which will be obtained by XFELs. The method is based on the maximum a
posteriori (MAP) estimation of the Bayesian statistics. The posterior
distribution is computed from the likelihood and the prior. For the
present problem, the Poisson distribution is appropriate for the
likelihood. The prior reflects the assumption of the electron
density. Since the electron density is known to have a lot of zeros,
we employ the exponential distribution which encourages the MAP
estimate to have a lot of zeros (this is called a sparse
solution~\cite{Tibshirani1996jrssB,ZhaoYu2006jmlr}. A similar idea is
employed in~\cite{Marchesini2008arxiv}). Therefore, we call our
method, sparse phase retrieval (SPR) method.

This combination is effective for the problem and is distinct from
other Bayesian approaches proposed for different phase retrieval
problems~\cite{IrwanLane1998josaa,BaskaranMillane1999ieeeip}. Compared
to the widely used hybrid input-output (HIO)
method~\cite{Fienup1978optlett,Fienup1982appopt,MiaoSayreChapman1998josaa},
the SPR method does not require the support region and is robust
against the limited photon counts and the missing central pixels. In
this paper, we focus on 2D phase retrieval problem. The resulting SPR
method can be used directly for the second scenario. If 3D diffraction
data are available, it is also possible to apply the SPR method. But
reconstruction of 3D diffraction data from 2D images with unknown
orientations is beyond the scope of the present paper.

\section{Proposed method}
\label{sec:theory}

\subsection{Diffraction image and noise}
\label{subsec:observation}

We first formulate the phase retrieval problem of a single diffraction
image. Let $f(x,y)\ge 0$ be the electron density of a molecule
projected onto a two-dimensional (2D) plane. The coordinate is
discretized, that is, $x,y= 1,\cdots,M$. We introduce the following
notation
\begin{equation*}
\V{f}=\{f(x,y)\}=\{f_{xy}\}=\{f_{11},\cdots,f_{MM}\},\hspace{1em}
\V{F}=\{F(u,v)\}=\{F_{uv}\}=\{F_{11},\cdots,F_{MM}\},
\end{equation*} where, $\V{F}$ is the Fourier transform of $\V{f}$
defined as follows,
\begin{equation}
  \label{eq:Fourier} F_{uv} = (\mathcal{F}(\V{f}))_{uv} = \frac{1}{M}
\sum_{x,y} f_{xy}\exp\Bigl(\frac{2\pi i(ux+vy)}{M}\Bigr).
\end{equation} The coefficient of the Fourier transform differs from
the commonly used $1/M^2$. We prefer the above definition because the
power of $\V{F}$ and $\V{f}$ is the same. We assume that the frame
size is sufficiently larger than the molecule in the image and many
components of $\V{f}$ are zero.

In the ideal situation, each pixel of a diffraction image receives the
number of photons proportional to the power spectrum $|F_{uv}|^2$,
where $(u,v)$ is the 2D index of the pixel. However, the XFEL
measurement of a single molecule diffraction will be different because
the flux of the x-ray laser is not sufficiently large. The number of
the photons detected by each sensor will be small and behave
stochastically.

\begin{figure}[t]
  \centering 
  \includegraphics[width=.65\figurewidth]{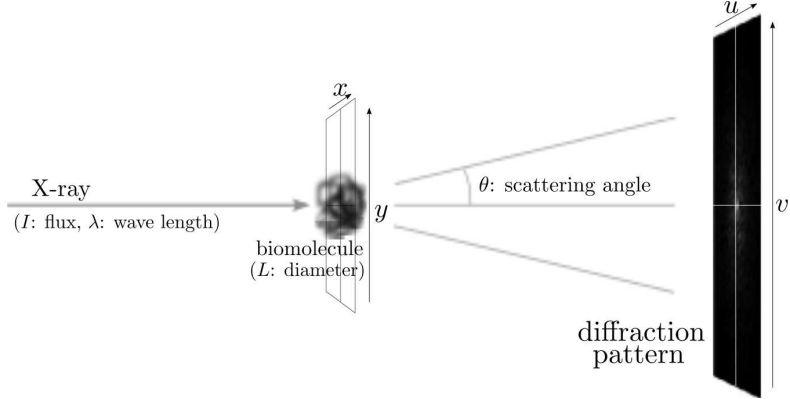}
  \caption{\label{fig:schematic}A schematic picture of the single
    biomolecular diffraction imaging with XFEL.
  }
\end{figure}

Let $N_{uv}$ be the number of photons detected by the sensor at
$(u,v)$. We study the distribution of $N_{uv}$. The total number
$N_{all}=\sum_{uv}N_{uv}$ is a stochastic variable which follows the
Poisson distribution. The expected value of $N_{all}$ is proportional
to the intensity $I_X$ of x-ray. Assuming each $N_{uv}$ is independent
and the distribution belongs to the same family, $N_{uv}$ also follows
a Poisson distribution. Let the expected value of $N_{uv}$ be $S_{uv}$
and the distribution of $N_{uv}$ becomes
\begin{equation*} 
  p(N_{uv}|S_{uv}) = \frac{S_{uv}^{N_{uv}}\exp(-S_{uv})}{N_{uv}!}.
\end{equation*}
We only consider this counting process and do not consider other
observation noise. Approximately, $S_{uv}$ is denoted as
$S_{uv}=\alpha|F_{uv}|^2\cos^3\theta$, where $\alpha$ is a positive
constant and $\theta$ is the scattering angle (see
Fig.~\ref{fig:schematic}). Because $\theta$ is a function of $u$ and
$v$, we further rewrite it as $\alpha |F_{uv}|^2c_{uv}$, where
$c_{uv}=c(u,v)=\cos^3\theta$.

The Fourier transform, $\V{F}$, is a linear function of $\V{f}$, and
the probability of $\V{N}$, where
$\V{N}=\{N_{uv}\}=\{N_{11},\cdots,N_{MM}\}$, is denoted as follows
\begin{equation}
  \label{eq:likelihood} 
  p(\V{N}|\V{f}) = \prod_{uv}
  \frac{(|F_{uv}|^2c_{uv})^{N_{uv}}\exp(-|F_{uv}|^2c_{uv})}{N_{uv}!}.
\end{equation} 
Note that the scaling of $|F_{uv}|^2$ cannot be determined only from
the observed scattering pattern and $\alpha$ is set to $1$. This
scaling indeterminacy is ignored here. It would be recovered from
other knowledge, such as the total energy.

In the following, we consider the density reconstruction as an
estimation of $\V{f}$. Commonly used estimate is the maximum
likelihood estimate (MLE) which maximizes $p(\V{N}|\V{f})$. The
likelihood in Eq.\eqref{eq:likelihood} is maximized by setting
$|F_{uv}|^2=N_{uv}/c_{uv}$, but this is not sufficient for the
estimation of $\V{f}$ because the phase is lost. Some additional
information is needed.

\subsection{SPR method}
\label{subsec:Bayesian Sparse}

Instead of the MLE, the MAP estimate of the Bayesian statistics is
used for the SPR method. The MAP estimate maximizes the posterior
distribution, where the posterior distribution is defined as the
product of the likelihood function and the prior distribution. Similar
framework has been proposed for different phase retrieval
problems~\cite{IrwanLane1998josaa,BaskaranMillane1999ieeeip}.

The likelihood function is defined as a Poisson distribution as in
Eq.\eqref{eq:likelihood} and we discuss the prior distribution in this
subsection. For a 2D image of a single molecule diffraction, it is
natural to assume that the electron density is high in the central
part and becomes zero in the peripheral part. But we do not know which
pixel is zero. The prior should reflect this knowledge.

\begin{figure}[tbp]
  \centering 
  \includegraphics[width=.5\figurewidth]{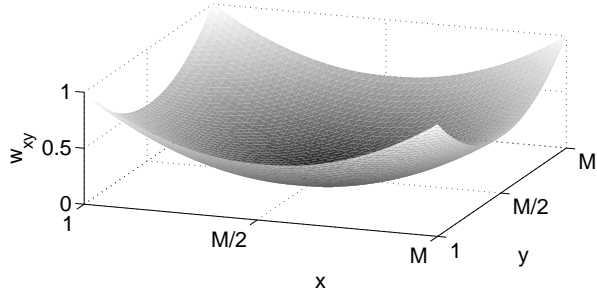}
  \caption{\label{fig:prior}The 3D plot of $w_{xy}$. This reflects the
    prior knowledge of the electron density.}
\end{figure}

It is known that the prior distribution with
an exponential distribution encourages the MAP estimate to have a lot
of zeros~\cite{Tibshirani1996jrssB,ZhaoYu2006jmlr} and we use this for
the SPR method
\begin{equation*} 
  p(\V{f}) =
  \prod_{xy}p(f_{xy}),\hspace{1em}
  p(f_{xy})=\rho_{xy}\exp(-\rho_{xy} f_{xy}),
  \hspace{2em} f_{xy}\ge 0,~~\rho_{xy}\ge 0,
\end{equation*} 
where each $f_{xy}$ is assumed to be independent. The hyperparameter
$\rho_{xy}$ reflects the prior belief of $f_{xy}=0$. Larger the value
of $\rho_{xy}$, stronger the belief. In order to reflect the prior
knowledge, we define the following one parameter function
$\rho(\mu)_{xy}$
\begin{equation}
  \label{eq:rho}
  \rho(\mu)_{xy} = \mu w_{xy},
  \hspace{1em}
    w_{xy}=
    a
    \biggl\{
    \Bigl(x-\frac{1+M}{2}\Bigr)^2+
    \Bigl(y-\frac{1+M}{2}\Bigr)^2
    \biggr\}+b.
\end{equation}
The two parameters $a$ and $b$ were adjusted to make $w_{xy}$ equal to
$0$ at the center and 1 at the corners (see Fig.~\ref{fig:prior}) and
$\mu$ adjusts the overall strength of the prior. We have employed a
parabolic function, but if there is any prior knowledge,
$\rho(\mu)_{xy}$ should reflect it.

The posterior distribution of $\V{f}$ observing $\V{N}$ has the
following relation,
\begin{equation*}
  p(\V{f}|\V{N})\propto p(\V{N}|\V{f})p(\V{f}).
\end{equation*}
It is convenient to take the logarithm of $p(\V{f}|\V{N})$ in order to
compute the MAP estimate. By taking the logarithm and collecting terms
related to $\V{f}$, the following function is obtained
\begin{equation}
  \label{eq:proposed loss}
  \ell_\mu(\V{f}|\V{N}) =\sum_{uv} \bigl(N_{uv} \ln
  |F_{uv}|^2-|F_{uv}|^2c_{uv}\bigr)
  -\mu\sum_{xy}w_{xy}f_{xy}.
\end{equation}
The first summation corresponds to the likelihood with the Poisson
distribution (similar idea is used in \cite{IrwanLane1998josaa}), the
second summation represents the prior with the exponential
distribution, and $\mu$ controls the balance between two terms. This
combination is the key idea of the proposed method. The SPR method is
summarized as follows.
\begin{align}
  \label{eq:SPR method}
  \mbox{maximize~~}\ell_\mu(\V{f}|\V{N})
  \hspace{1em}
  {\mbox{subject to}} \hspace{1em} f_{xy}\ge 0 \hspace{1em}(1\le
  x,y\le M).
\end{align}
We start with a large $\mu$ and decrease it. The MAP estimate is
computed for each $\mu$, and the best value of $\mu$ is selected
according to some criterion. This issue will be discussed later.

We note that our formulation is related to the compressed sensing
(CS)~\cite{Donoho2006ieeeit}, and the idea of CS has been applied to
the phase retrieval problem~\cite{Marchesini2008arxiv}.  Our approach
is different from the work since the Poisson likelihood is included.

\subsection{Algorithm}
\label{subsec:algorithm}

A gradient based algorithm is used to compute the MAP estimate. The
algorithm is shown below starting with the following relation,
\begin{equation*}
  \frac{\partial |F_{uv}|^2}{\partial f_{xy}}
  =\frac{2}{M}
  \mathrm{Re}
  \biggl(
  F_{uv}
  \exp
  \Bigl(-\frac{2\pi i(ux+vy)}{M}\Bigr)
  \biggr).
\end{equation*}
By defining the inverse Fourier transform as 
\begin{equation*}
  (\mathcal{F}^{-1}(\V{F}))_{xy}
  =
  \frac{1}{M}
  \sum_{uv}
  F_{uv}\exp\Bigl(-\frac{2\pi i(ux+vy)}{M}\Bigr),
\end{equation*}
the derivative of $\ell_\mu(\V{f}|\V{N})$ becomes
\begin{equation*}
  \frac{\partial \ell_\mu(\V{f}|\V{N})}{\partial f_{xy}}
  =
  2\mathrm{Re}\bigl(\mathcal{F}^{-1}(\V{g}(\V{F};\V{N}))\bigr)_{xy}
  -\mu w_{xy},
\end{equation*}
where $\V{g}(\V{F};\V{N})=\{g_{uv}(F_{uv};N_{uv})\}$ and 
\begin{equation*}
  g_{uv}(F_{uv};N_{uv})=
  \biggl(
  \frac{N_{uv}}{|F_{uv}|^2}
  -c_{uv}
  \biggr)F_{uv}.
\end{equation*}
In order to compute the derivative of $\ell_\mu(\V{f}|\V{N})$,
$\mathcal{F}$ is applied to $\V{f}$, then $\mathcal{F}^{-1}$ is
applied to $\V{g}(\V{F};\V{N})$. Thus, the computational cost is
similar to the HIO method~\cite{Fienup1982appopt}.

We employed a na\"ive iterative updating rule for the MAP estimate of
$f_{xy}$, $x,y=1,\cdots,M$.
\begin{equation}
  \label{eq:updating rule}
  f^{t+1}_{xy}=
  \max\Bigl(0,\ f^{t}_{xy}+
  \eta_t\frac{\partial \ell_\mu(\V{f}^{t}|\V{N})}{\partial
    f^{t}_{xy}}\Bigr),
\end{equation}
where $t$ is the iteration number. The ``max'' operation keeps
$f_{xy}$ nonnegative. The positive variable $\eta_t$ controls the step
size and a simple line search of $\eta_t$ accelerated the convergence
largely.

The initial value is generally important for gradient based
algorithms. But for the SPR method, the influence is effectively
reduced by controlling $\mu$. The MAP estimate is first computed for a
large $\mu$ and then for smaller $\mu$'s successively. For a very
large $\mu$, most components of the optimal $\V{f}$ are 0 and the
influence of the initial value is negligible. The MAP estimate is then
used for the initial value of a slightly smaller $\mu$, which is a
good starting point in general.

\subsection{HIO method and Bayesian framework}

In this subsection, we discuss the relation between the widely used HIO
method and the SPR method.

The HIO
method~\cite{Fienup1978optlett,Fienup1982appopt,MiaoSayreChapman1998josaa}
exploits the knowledge that the positive density is localized in a
small support region. Let us define the support region $\gamma$, and
the method is written as follows,
\begin{align}
  \label{eq:HIO Bayes}
  \mbox{minimize}\sum_{uv}(|F_{uv}|c_{uv}^{1/2}-N_{uv}^{1/2})^2,
  \hspace{.5em}
  \mbox{subject to} \hspace{.5em}
  f_{xy}\ge 0, (x,y)\in \gamma \hspace{.5em}\mbox{and}\hspace{.5em} 
  f_{xy}=0,  (x,y)\notin \gamma.
\end{align}
This problem can be formulated as the MAP estimation problem by
setting the Gaussian likelihood function and the uniform prior
distribution, that is,
\begin{align*}
  p_{G}(\V{N}|\V{f})&=\frac{1}{(2\pi\sigma^2)^{M^2/2}}\exp
  \Bigl(
  -\frac{1}{2\sigma^2}\sum_{uv}(|F_{uv}|c_{uv}^{1/2}-N_{uv}^{1/2})^2
  \Bigr),~~
  \tilde{p}_{u}(f_{xy})=
  \left\{
    \begin{array}[h]{ll}
      C,& (x,y)\in \gamma\\
      0, & (x,y)\notin \gamma
    \end{array}
  \right.,
\end{align*}
where $C$ is a positive number. If $f_{xy}$ is unbounded, above
$\tilde{p}_{u}(f_{xy})$ is improper. The MAP estimate is computed by solving
Eq.\eqref{eq:HIO Bayes}.

Formally, the difference between the HIO method and the SPR method is
the likelihood and the prior. But more importantly, the proposed SPR
method does not require the support region. In the SPR method, a lot
of components of the estimated electron density $\V{f}$ become 0
automatically.

\section{Numerical Experiment}
\label{sec:Numerical Experiment}

\subsection{Simulated data}
\label{subsec:Simulated Data}

\begin{figure}[htbp]
  \centering 
  \subfloat[\label{fig:true}2D electron density of a lysozyme
  protein.]{
      \includegraphics[height=.4\figurewidth]{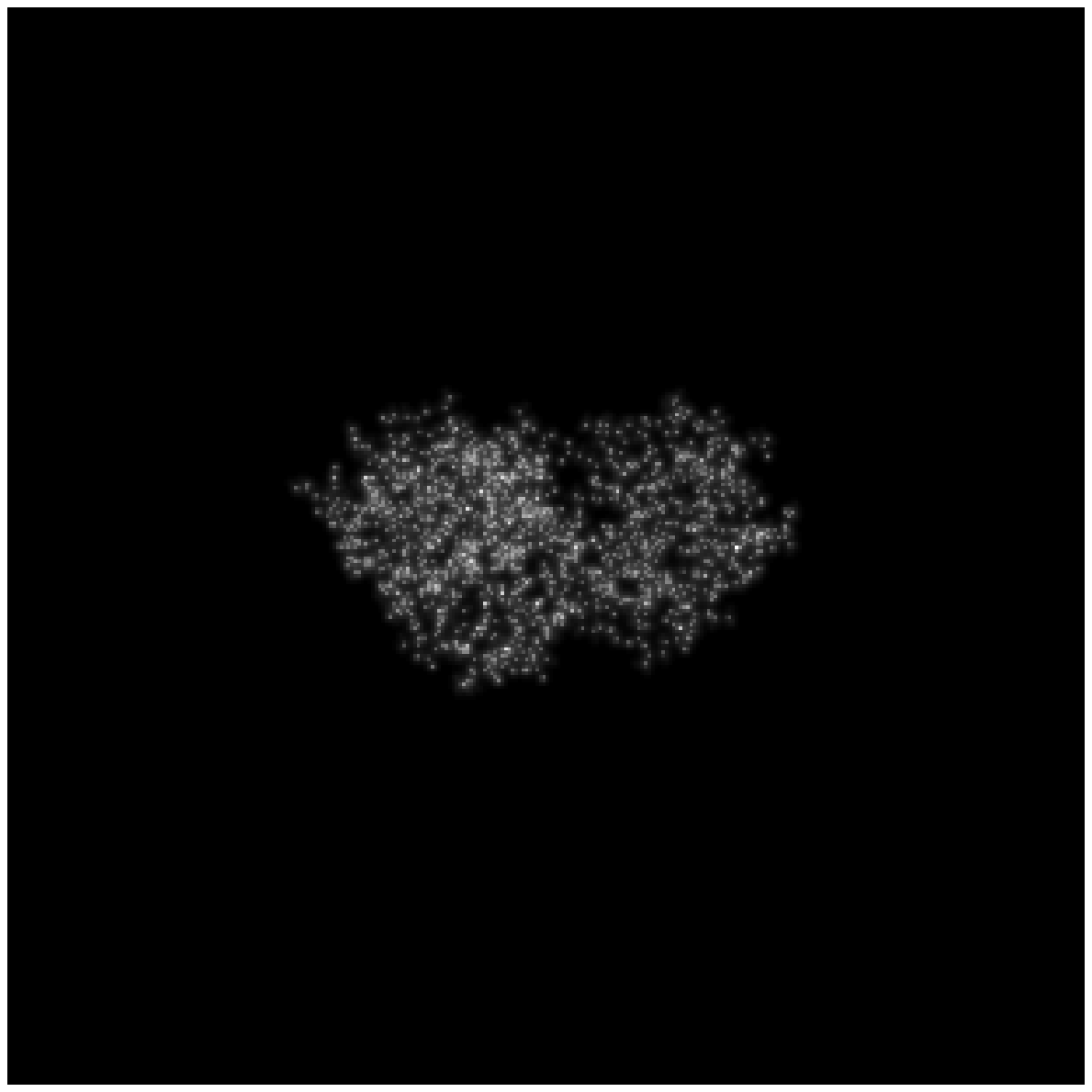}
    }
    \raisebox{-2.2pt}{
      \includegraphics[height=.418\figurewidth,bb = 470 238 536 596, clip]
      {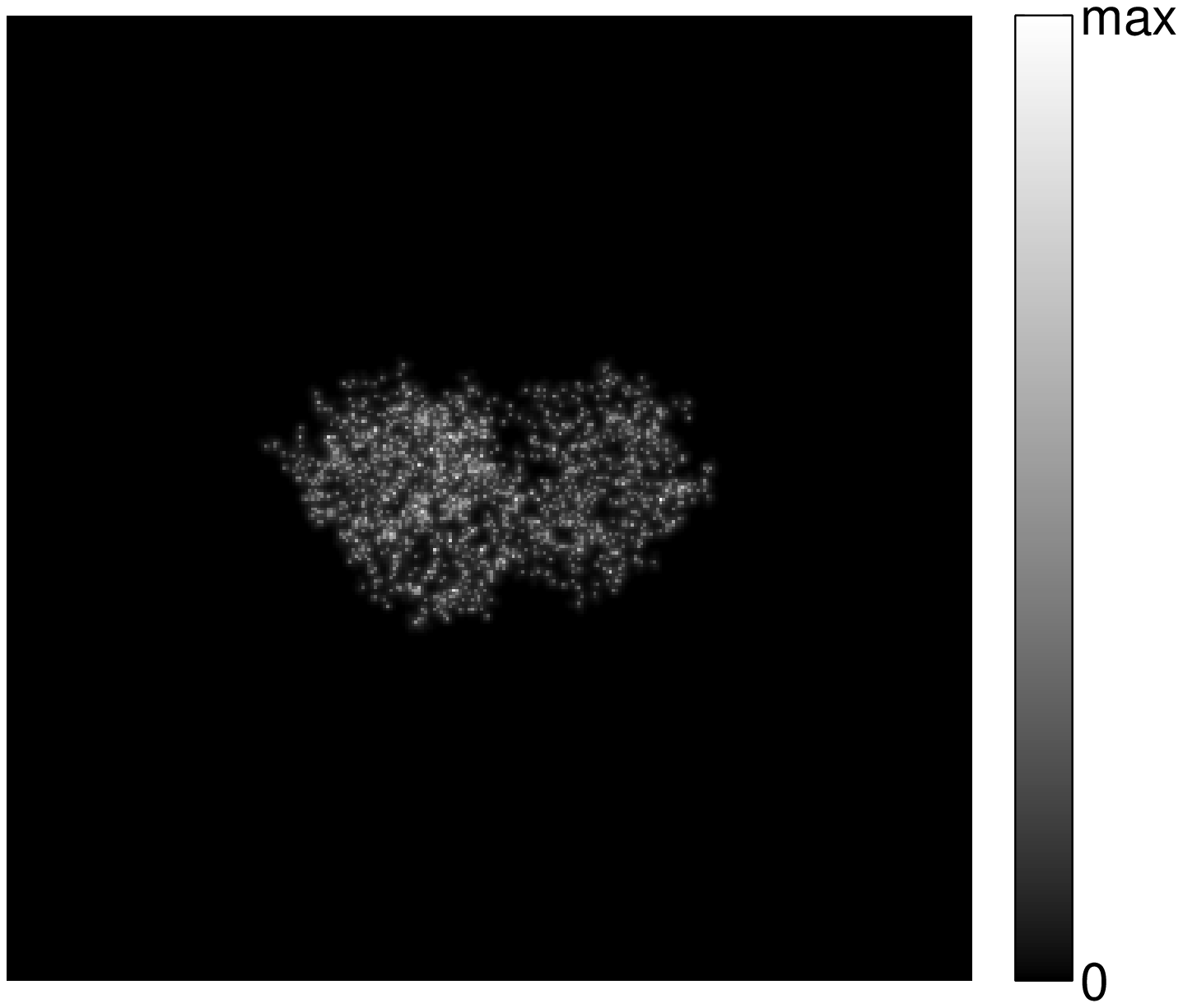}}\\
    \subfloat[\label{fig:true_diffraction} A noiseless diffraction
    image with x-ray fluxes $I=5.0\times 10^{22}$. The scale of the
    image is proportional to $\log(S_{uv}+1)$.]{
    \includegraphics[width=.27\figurewidth]{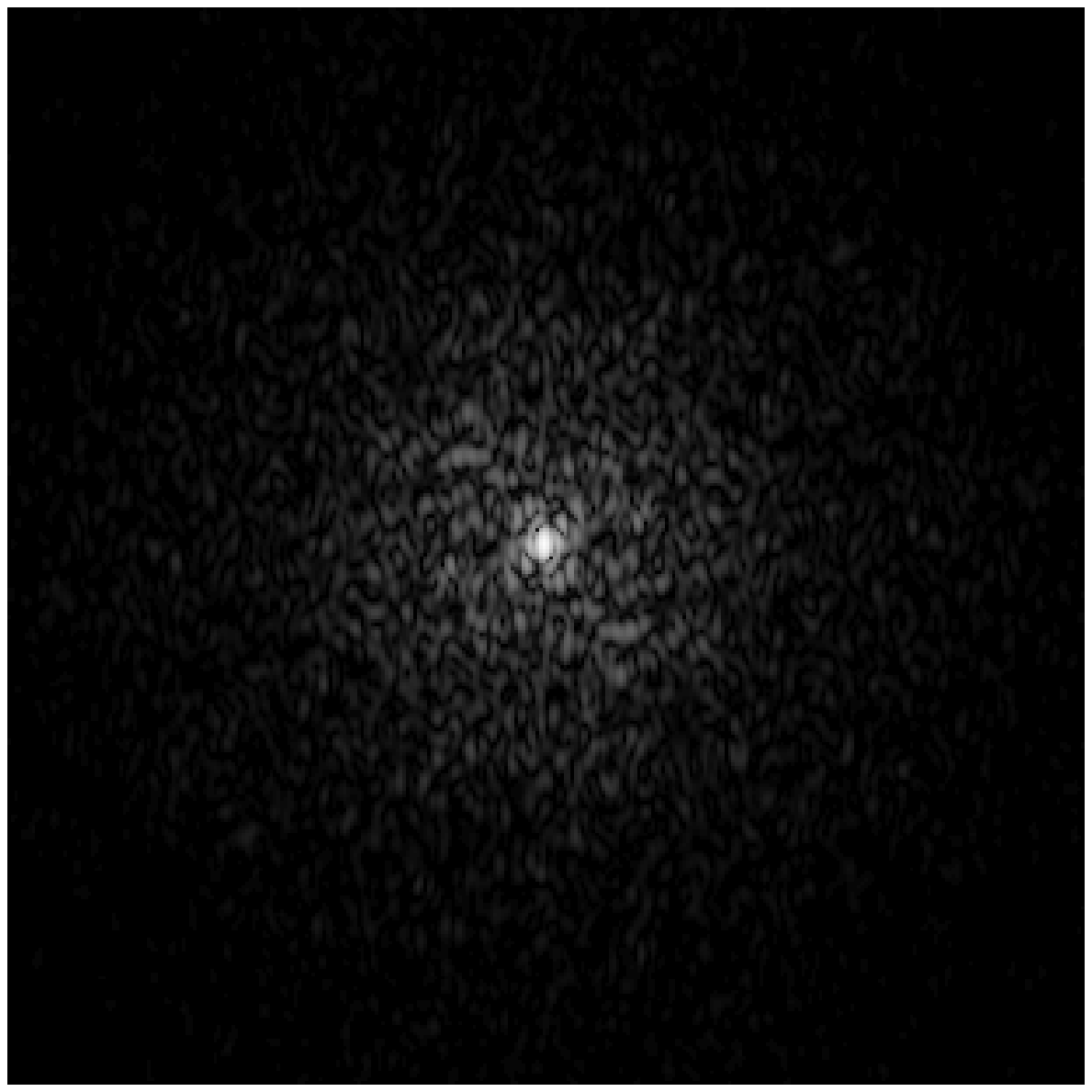}
  }\hfil
  \subfloat[\label{fig:noisy_diffraction_5e21}Simulated diffraction
  image with x-ray fluxes $I=5.0\times 10^{21}$. The total number of
  photons is $4630$. The scale
    of the image is proportional to $\log(N_{uv}+1)$.]{
    \includegraphics[width=.27\figurewidth]{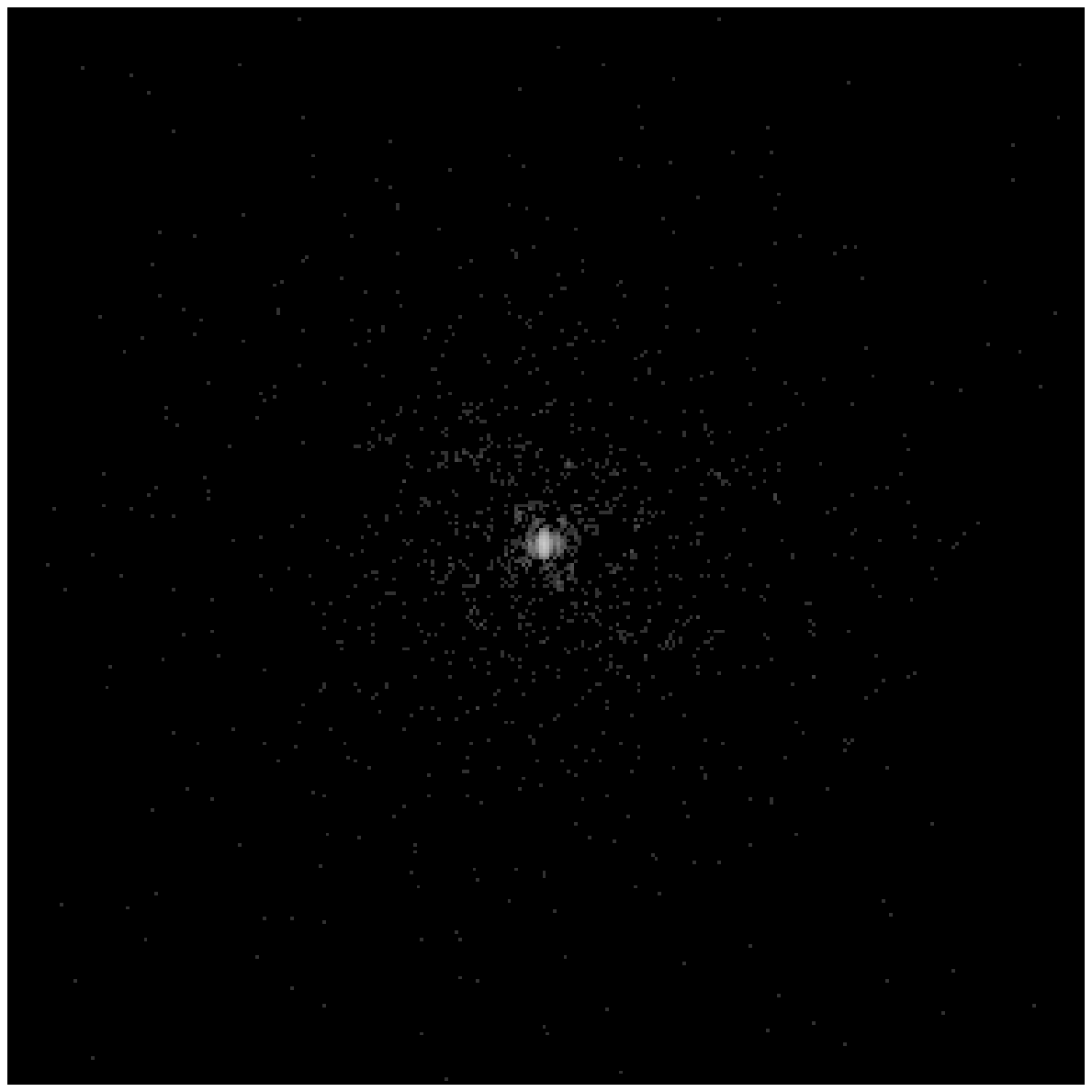}
  }\hfil
  \subfloat[\label{fig:noisy_diffraction_1e21}Simulated diffraction
  image with x-ray fluxes $I=1.0\times 10^{21}$. The total number of
  photons is $957$. The scale of the image is proportional to
  $\log(N_{uv}+1)$.]{
    \includegraphics[width=.27\figurewidth]{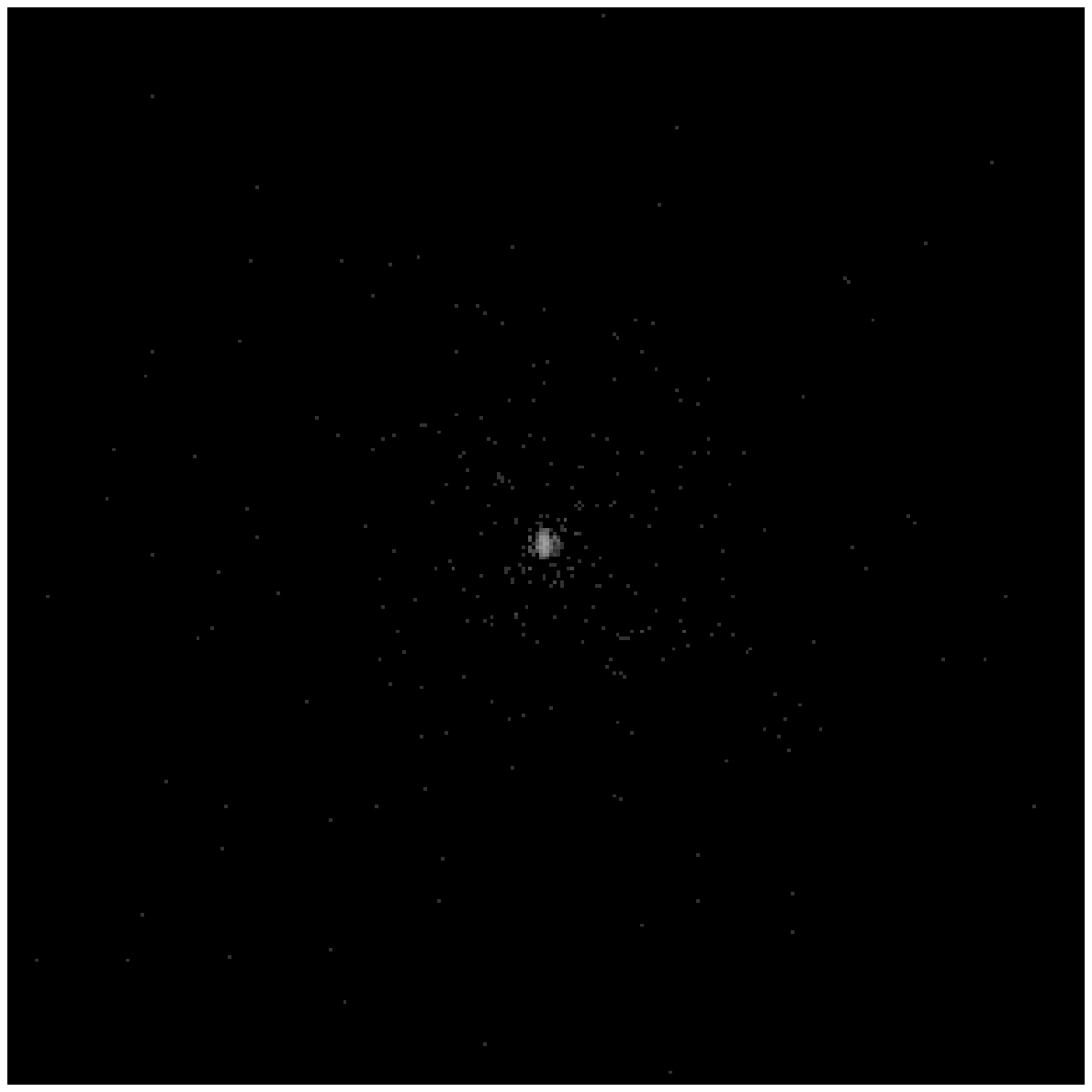}
  }
  \raisebox{-1.6pt}{
    \includegraphics[height=.282\figurewidth,bb = 470 238 536 596, clip]
    {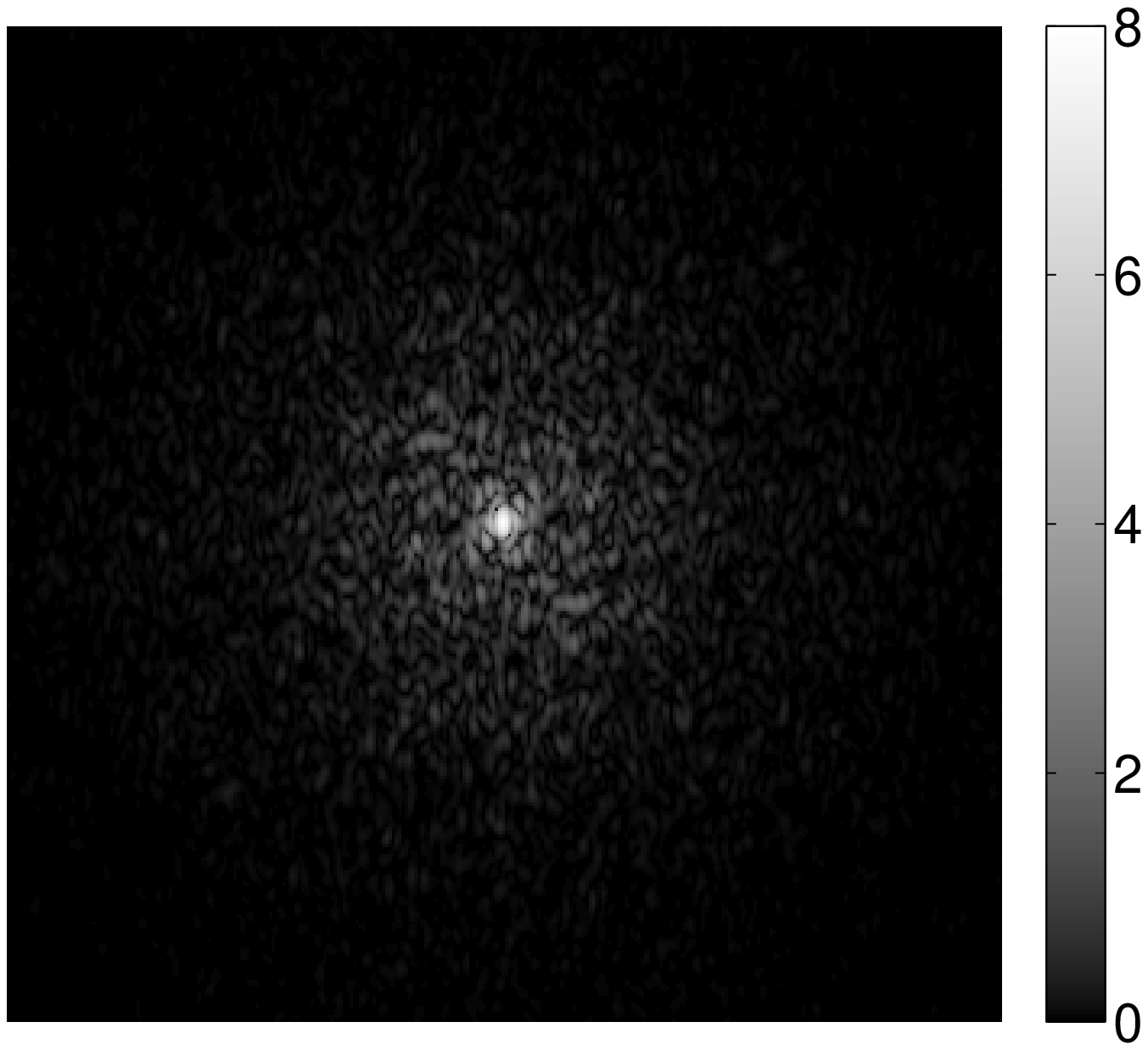}}
  \caption{\label{fig:lysozyme data}Simulated 2D electron density (a),
    the diffraction image without noise (b), and simulated diffraction
    patterns (c,d) of a lysozyme protein are shown. Each image is
    308$\times$308 pixels. The spatial frequency of the diffraction
    patterns is $0.0812~{\mathrm{nm}}^{-1}$ and the unit of the x-ray
    fluxes $I$ is $\photonsperpulsemm$.}
\end{figure}
The SPR method was tested with simulated diffraction data. The electron
density of lysozyme, a $14~\mathrm{kD}$ protein, was projected onto a
2D plane. The projected density was then Fourier transformed into
reciprocal space $(u,v)$ to simulate a diffraction
image. The density and the ideal diffraction image are shown in 
Figs.~\ref{fig:lysozyme data}a and \ref{fig:lysozyme data}b,
respectively.

Experimental diffraction images were simulated by converting the
diffraction intensity into the number of photons per effective pixel
$(\lambda/\sigma L)^2$ according to the Poisson distribution with the
mean, $S_{uv}$, defined as follows,
\begin{equation*}
  S_{uv}
  = \alpha |F_{uv}|^2c_{uv}
  =
  Ir_c^2
  \Bigr(
  \frac{\lambda}{\sigma L}
  \Bigl)^2
  |F_{uv}|^2c_{uv}.
\end{equation*}
Here, $I$ is the incident x-ray flux ($\photonsperpulsemm$), $r_c$ is
the classical electron radius ($2.82{\times}10^{-12}~\mathrm{mm}$),
$\lambda$ is the x-ray wave length ($0.1~\mathrm{nm}$), $\sigma$ is
the oversampling ratio per one dimension ($2$), and $L$ is the
molecule diameter ($6.16~\mathrm{nm}$) (see
Fig.~\ref{fig:schematic}). The Ewald sphere curvature was ignored but
this is a good approximation for small angles.

Two examples for scattered photons of x-ray fluxes with
$I=5.0{\times}10^{21}$ and $1.0{\times}10^{21}$ $\photonsperpulsemm$
are shown in Figs.~\ref{fig:lysozyme data}c and \ref{fig:lysozyme
  data}d, respectively. Although the fluxes in
the above simulation are around 100 times larger than that of the
current LCLS, the sizes of the molecules of interest are generally
larger than lysozyme. When the size is around $1000~\mathrm{kD}$, the
total number of photons arriving at the 2D plane will be in the
similar range as Figs.~\ref{fig:lysozyme data}c and 
\ref{fig:lysozyme data}d.

\subsection{Results}
\label{subsec:results}

\begin{figure}[htbp]
  \centering 
  \subfloat[Reconstructed electron density with $\mu=10000$.]{
    \includegraphics[width=.27\figurewidth]{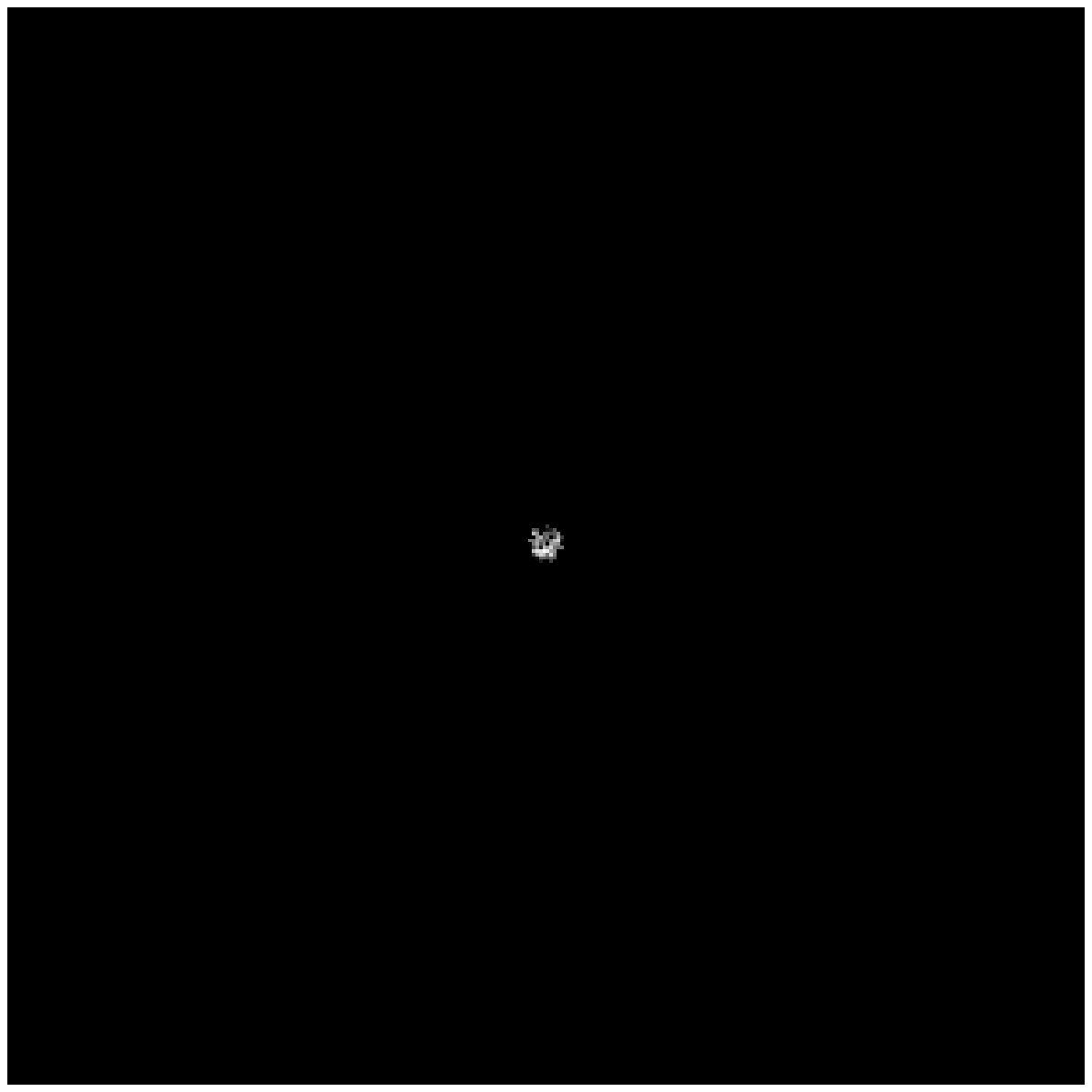}
  }\hfil
  \subfloat[Reconstructed electron density with $\mu=100$.]{
    \includegraphics[width=.27\figurewidth]{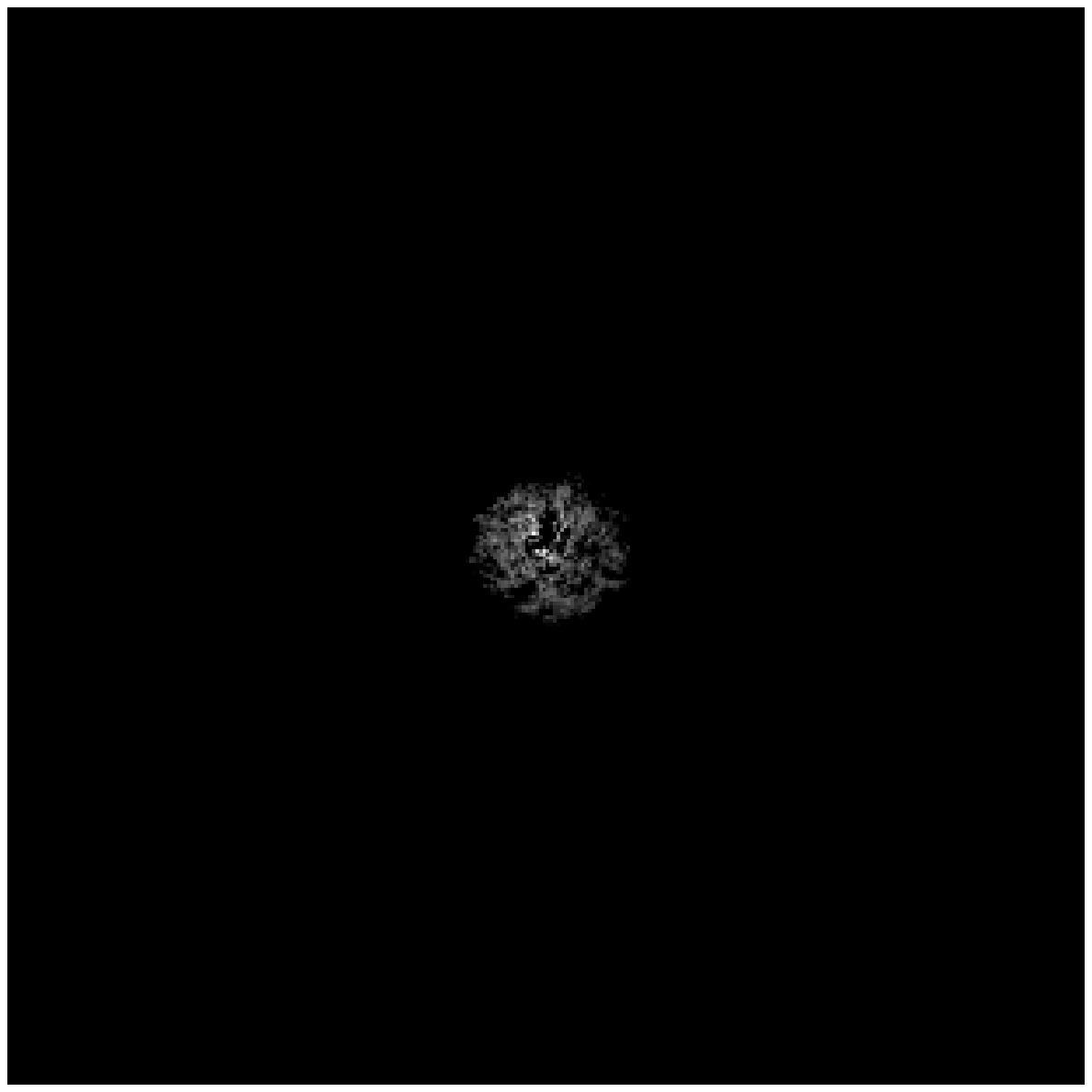}
  }\hfil
  \subfloat[Reconstructed electron density with $\mu=1$.]{
      \includegraphics[width=.27\figurewidth]{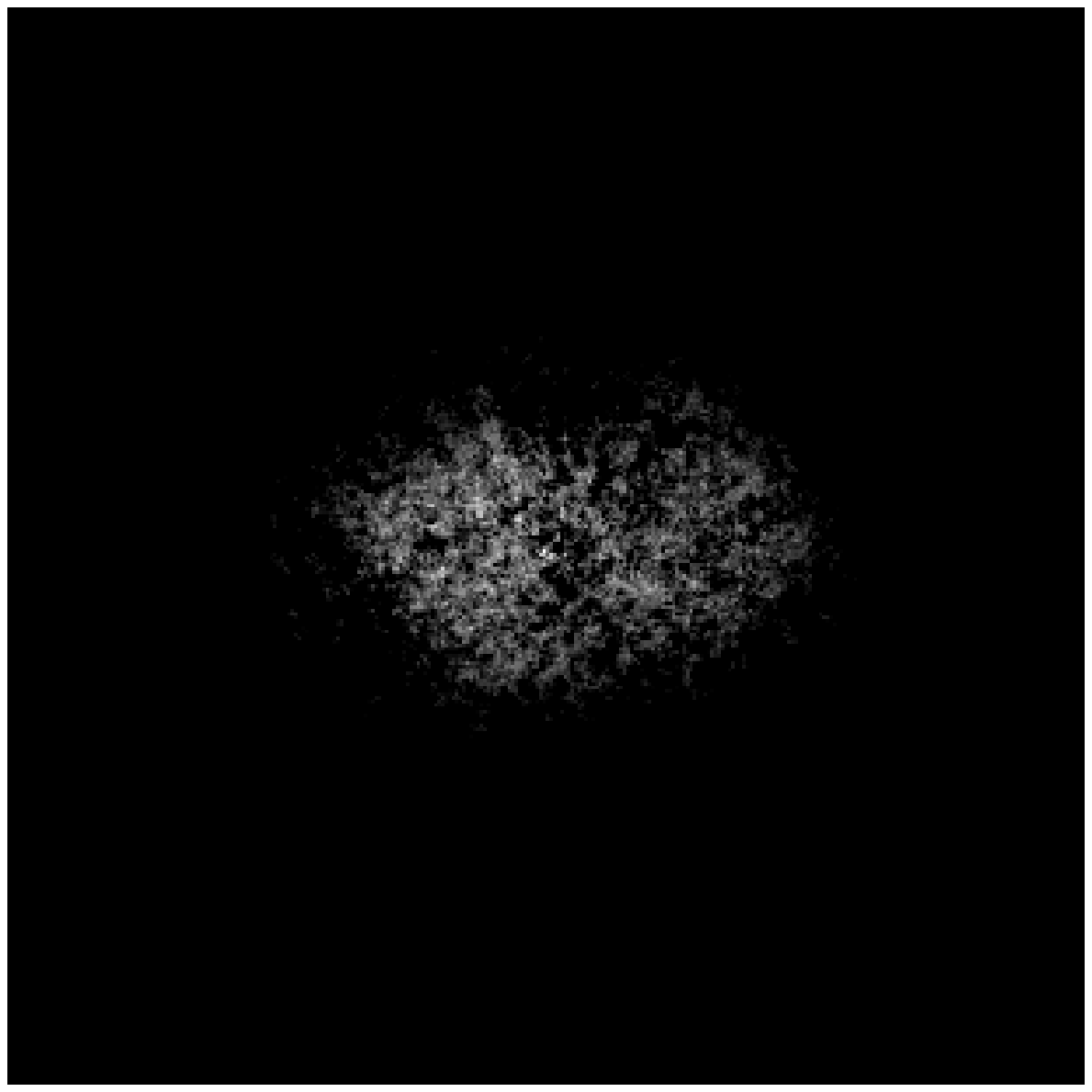}
    }
    \raisebox{-1.6pt}{
      \includegraphics[height=.282\figurewidth,bb = 470 238 536 596, clip]
      {colorbar.eps}}
    \caption{\label{fig:SPR true}Reconstruction of the electron
      density from the ideal diffraction image by the SPR method with
      different values of $\mu$.}
\end{figure}
The SPR method was applied to the data to reconstruct the density.
Figure \ref{fig:SPR true} shows the results of the SPR method applied
to the ideal diffraction image. As $\mu$ decreases, more components
become positive. The reconstructed electron density becomes similar to
the true density when $\mu=1$. Thus, it is important to select the
best $\mu$. A simple strategy is to rank the results with a
criterion. Many criteria have been proposed in statistics, such as,
Akaike Information Criterion (AIC) and Bayesian Information Criterion
(BIC)~\cite{ZouHastieTibshirani2007as}, but here we use a different
one. Let $\hat{\V{f}}(\mu)$ be the maximizer of
$\ell_\mu(\V{f}|\V{N})$ and the following error is used as the
criterion
\begin{equation}
  \label{eq:error_k}
  \mathrm{Error}_{F}(\mu) =
  \frac{\sum_{uv}(|\hat{F}(\mu)_{uv}|{c_{uv}}^{1/2}-
    {N_{uv}}^{1/2})^2}{\sum_{u'v'}N_{u'v'}},
\end{equation}
where $\hat{F}(\mu)_{uv}=\mathcal{F}(\hat{\V{f}}(\mu))_{uv}$. This
error function has been widely used for the phase retrieval problem
and is employed here as well in order to rank the results. Roughly
speaking, the error decreases as $\mu$ shrinks since it controls the
balance between the prior and the likelihood. But the error does not
necessarily decrease monotonically because the log posterior in
Eq.~\eqref{eq:proposed loss} is different from the error. Thus, we
choose the $\mu$ which minimizes $\mathrm{Error}_{F}(\mu)$.
\begin{figure}[tbp]
  \centering 
  \includegraphics[width=.75\figurewidth]{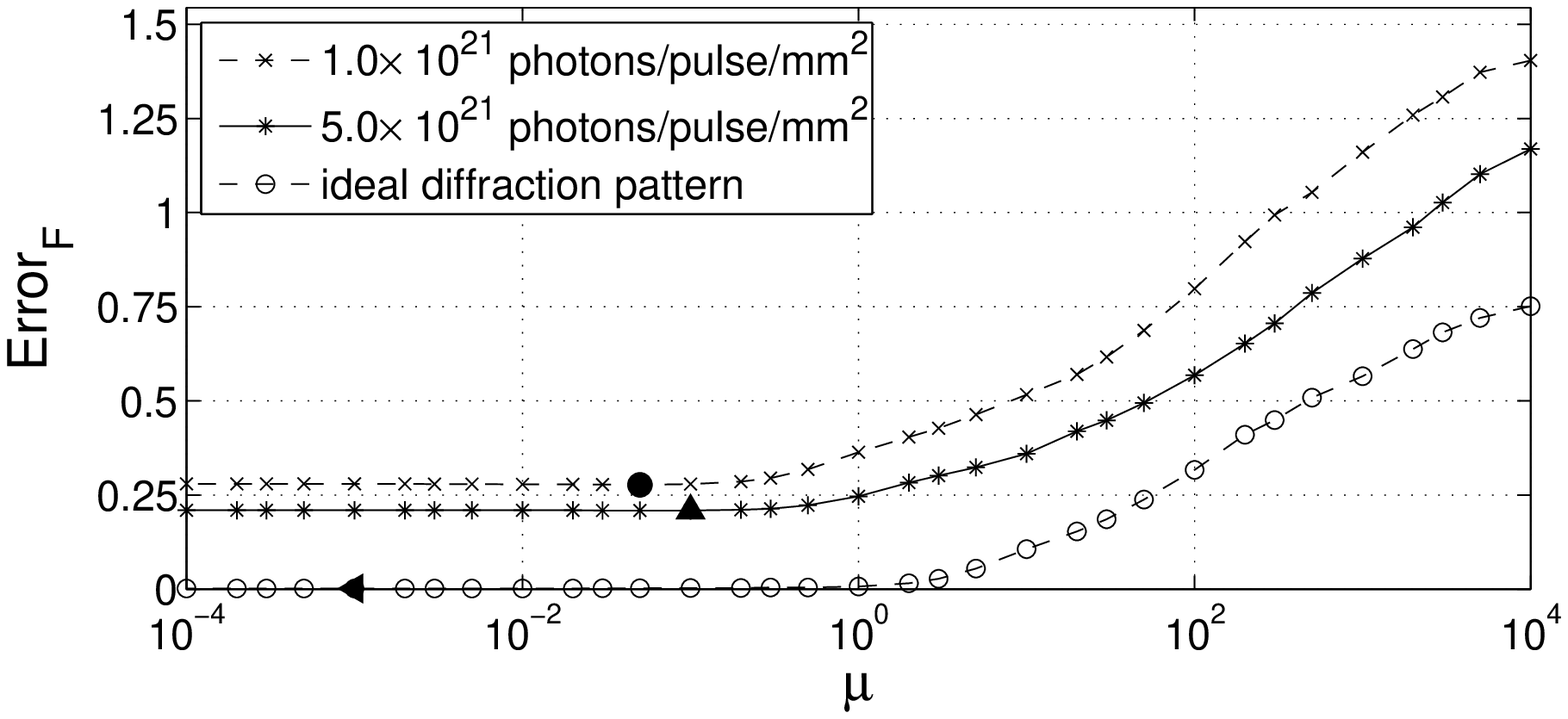}
  \caption{\label{fig:error_F}The values of
    $\mathrm{Error}_{F}(\mu)$ for simulated diffraction
    images in Fig.~\ref{fig:lysozyme data}. The black circle and
    triangles on the curves correspond to the reconstructed density
    images of Figs.~\ref{fig:SPR}.}
\end{figure}
\begin{figure}[tbp]
  \centering 
  \subfloat[Reconstruction from diffraction image Fig.2b by the SPR
  method with $\mu=0.001$. $\mathrm{Error}_{F}$ is
  1.60$\times10^{-3}$.]{
    \includegraphics[width=.27\figurewidth]{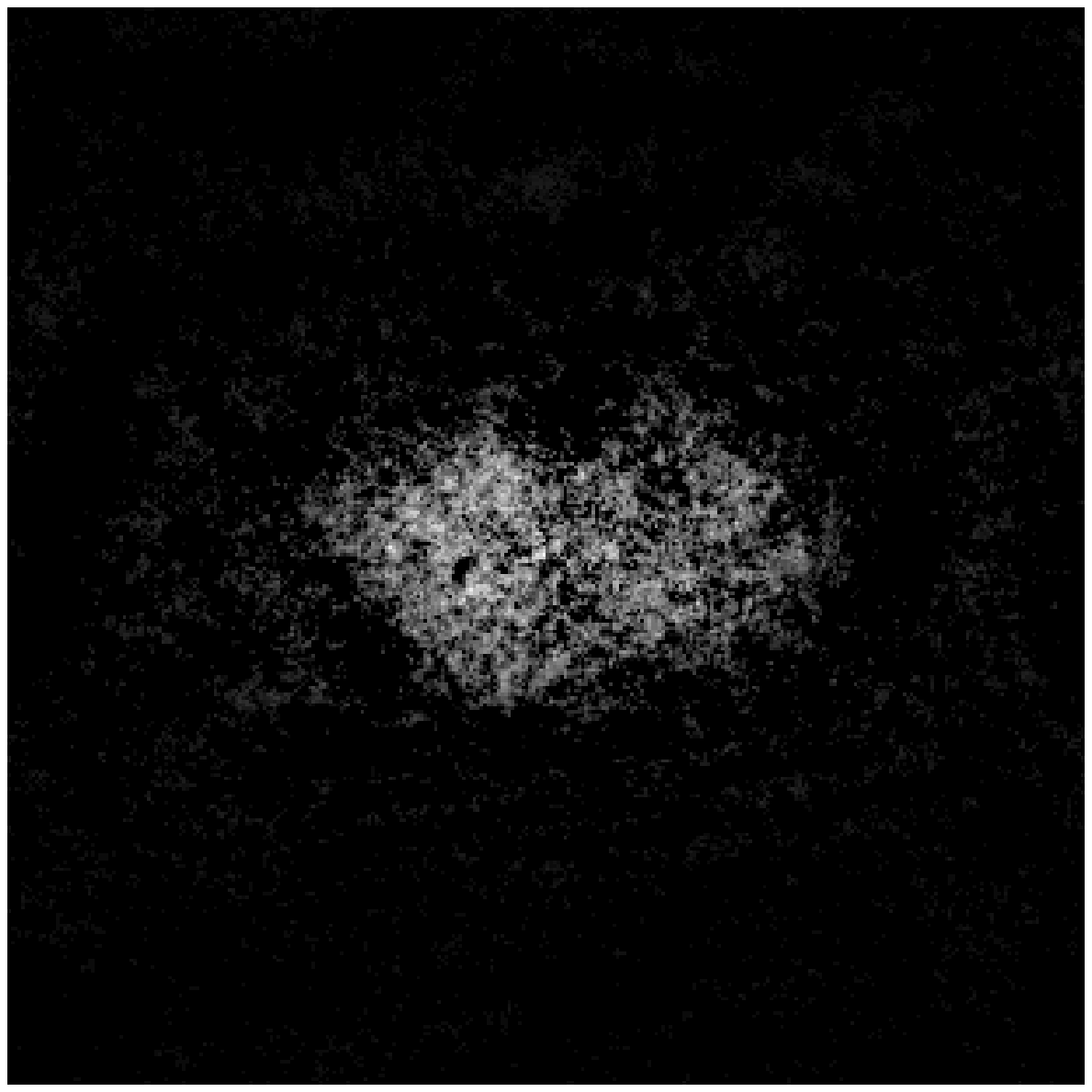}
  }\hfil
  \subfloat[Reconstruction from diffraction image Fig.2c by the SPR
  method with $\mu=0.1$. $\mathrm{Error}_{F}$ is 0.209.]{
    \includegraphics[width=.27\figurewidth]{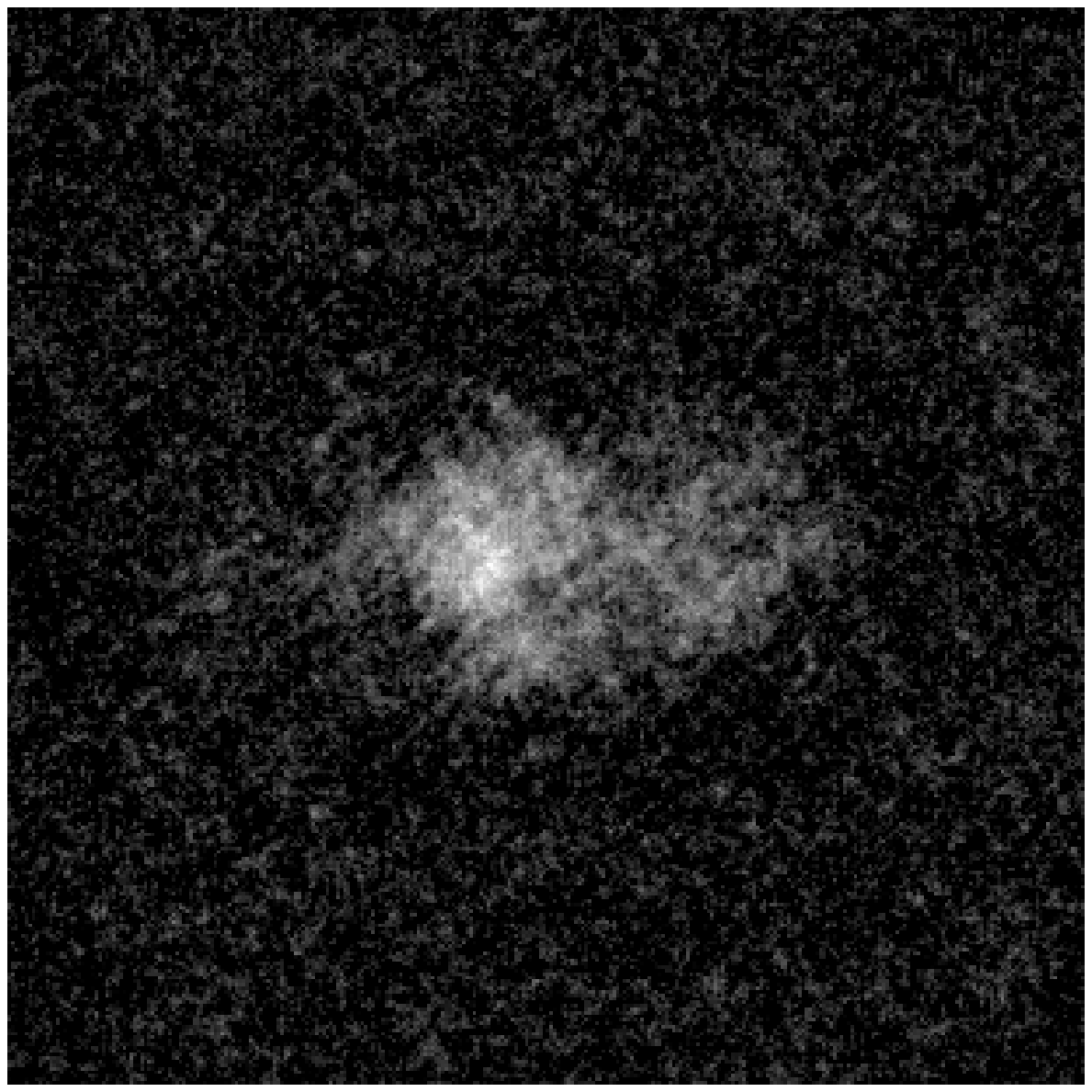}
  }\hfil
  \subfloat[Reconstruction from diffraction image Fig.2d by the SPR
  method with $\mu=0.05$. $\mathrm{Error}_{F}$ is 0.277.]{
    \includegraphics[width=.27\figurewidth]{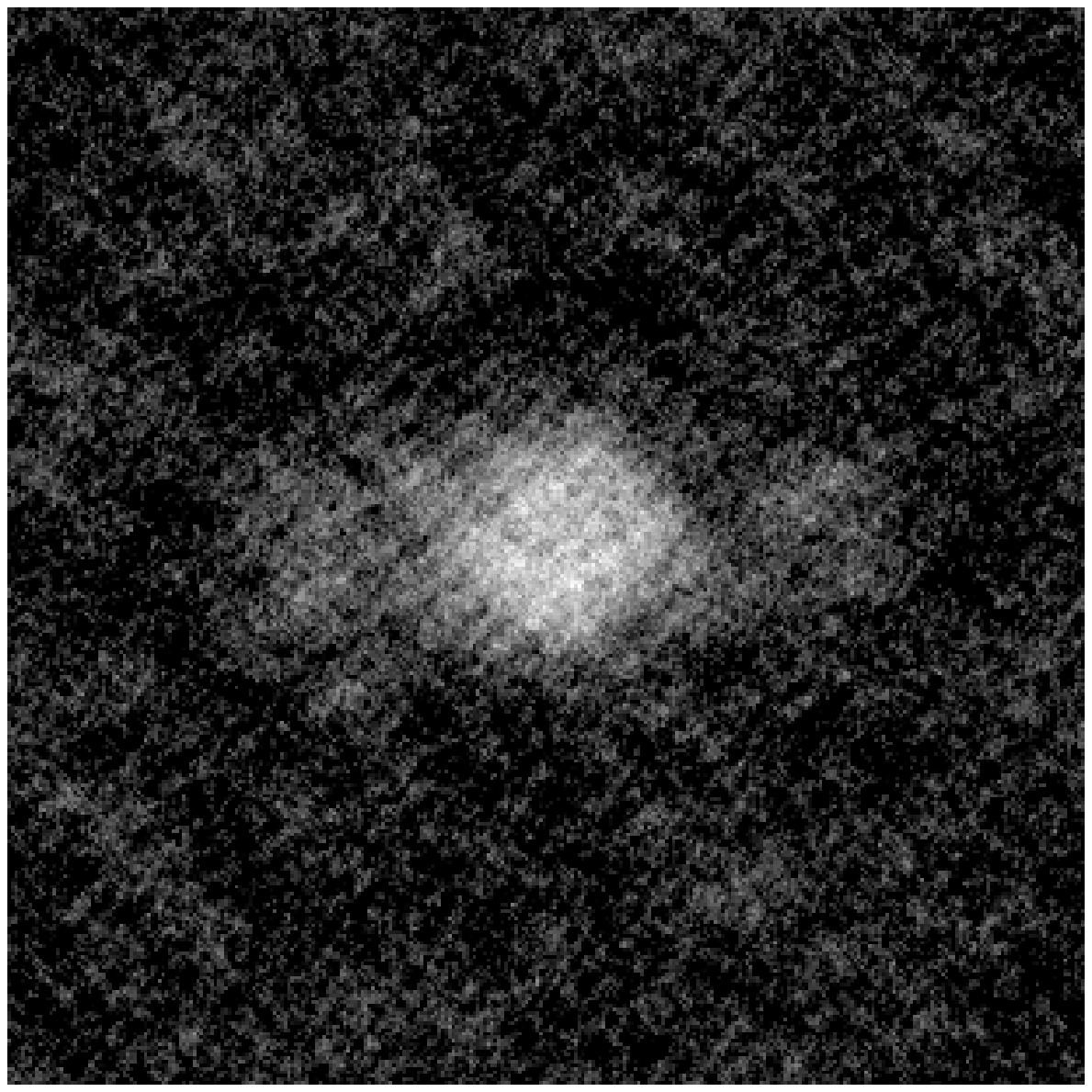}
  }
  \raisebox{-1.6pt}{
    \includegraphics[height=.282\figurewidth,bb = 470 238 536 596, clip]
    {colorbar.eps}}
  \caption{\label{fig:SPR}Reconstruction of the 2D electron density
    by the SPR method.}
\end{figure}
\begin{figure}[tbp]
  \centering 
  \subfloat[Reconstruction from diffraction image Fig.2b by the HIO
  method. $\mathrm{Error}_{F}$ is 3.30$\times10^{-3}$.]{
      \includegraphics[width=.27\figurewidth]{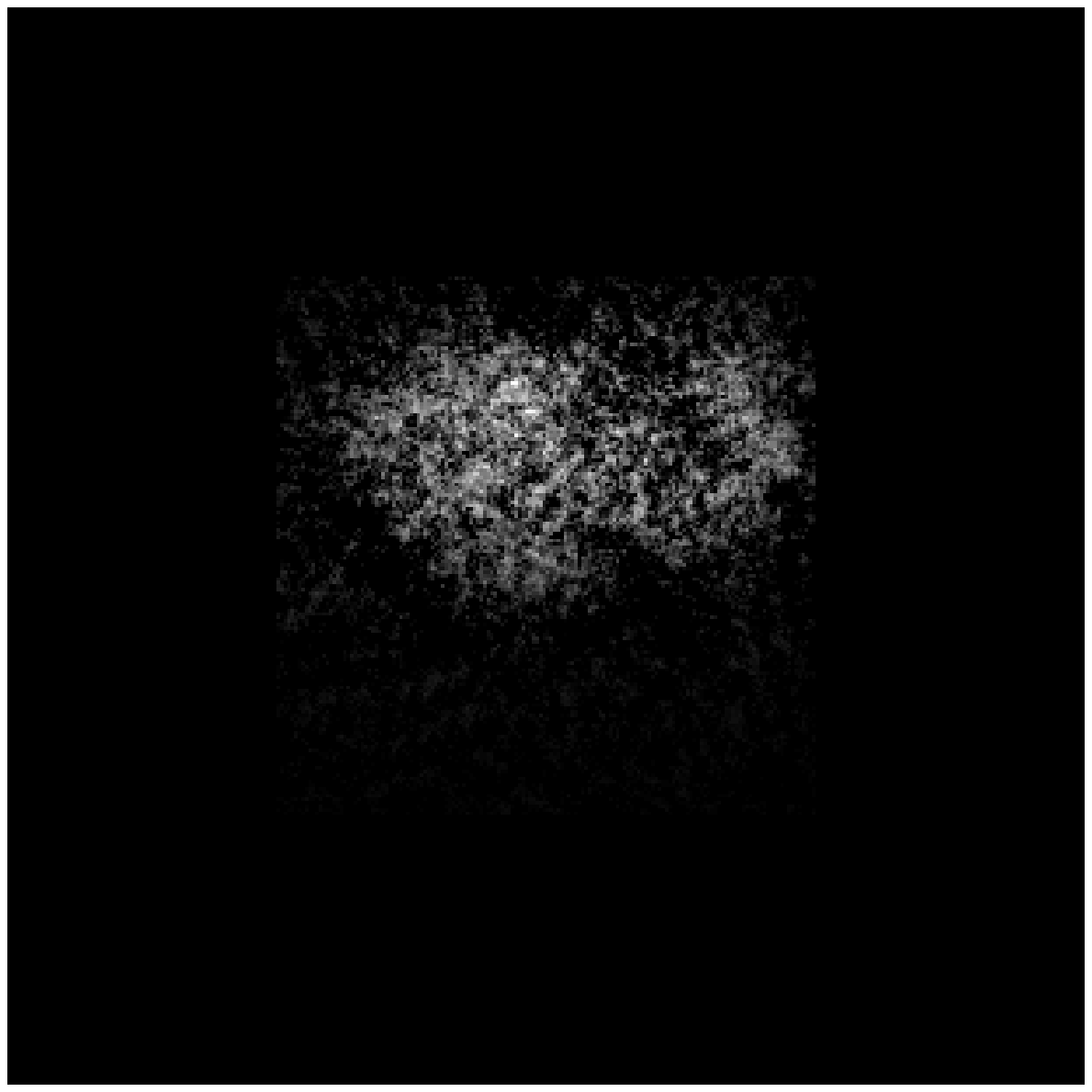}
    }\hfil
    \subfloat[Reconstruction from diffraction image Fig.2c by the HIO method.
  $\mathrm{Error}_{F}$ is 0.220.]{
      \includegraphics[width=.27\figurewidth]{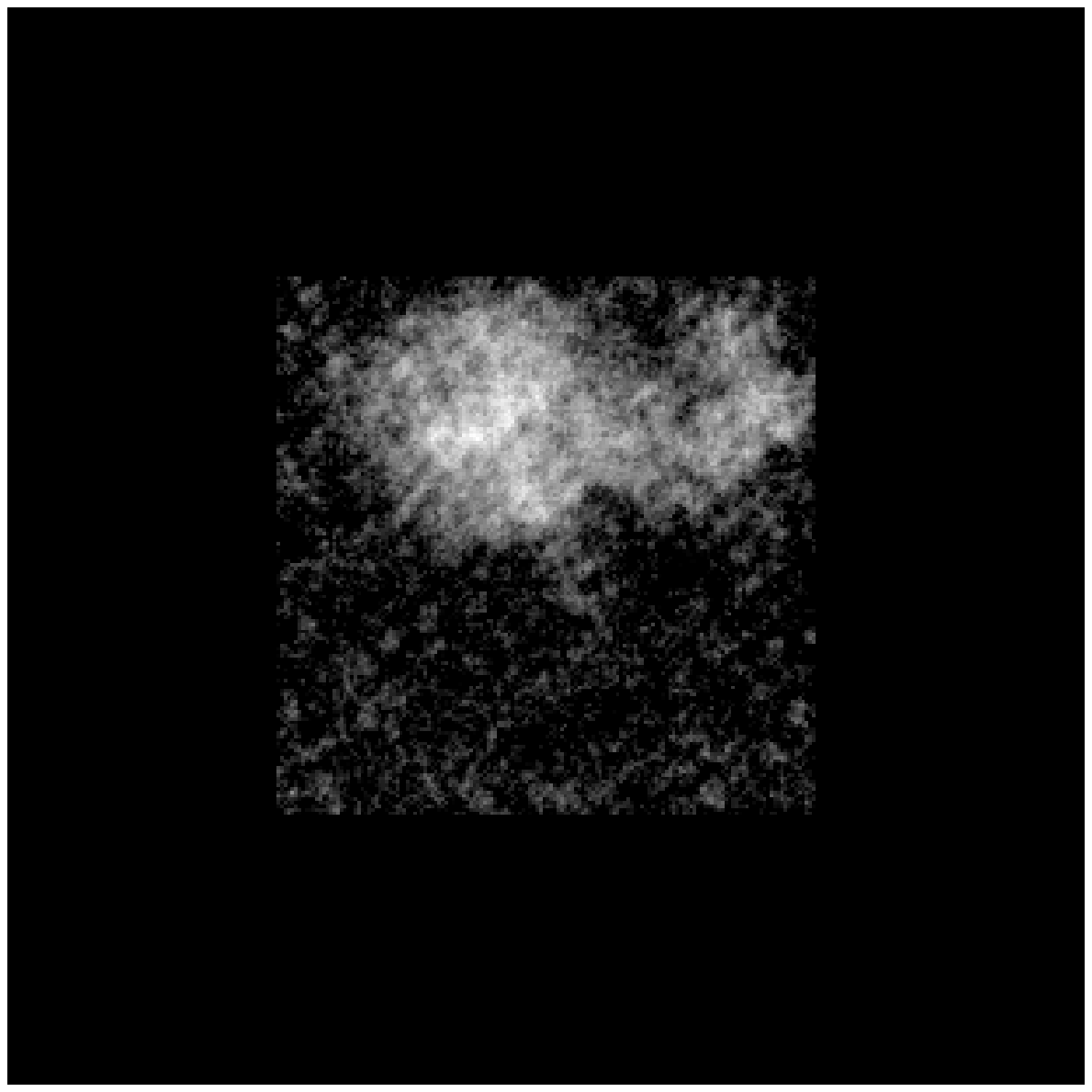}
    }\hfil
  \subfloat[Reconstruction from diffraction image Fig.2d by the HIO method.
  $\mathrm{Error}_{F}$ is 0.291.]{
    \includegraphics[width=.27\figurewidth]{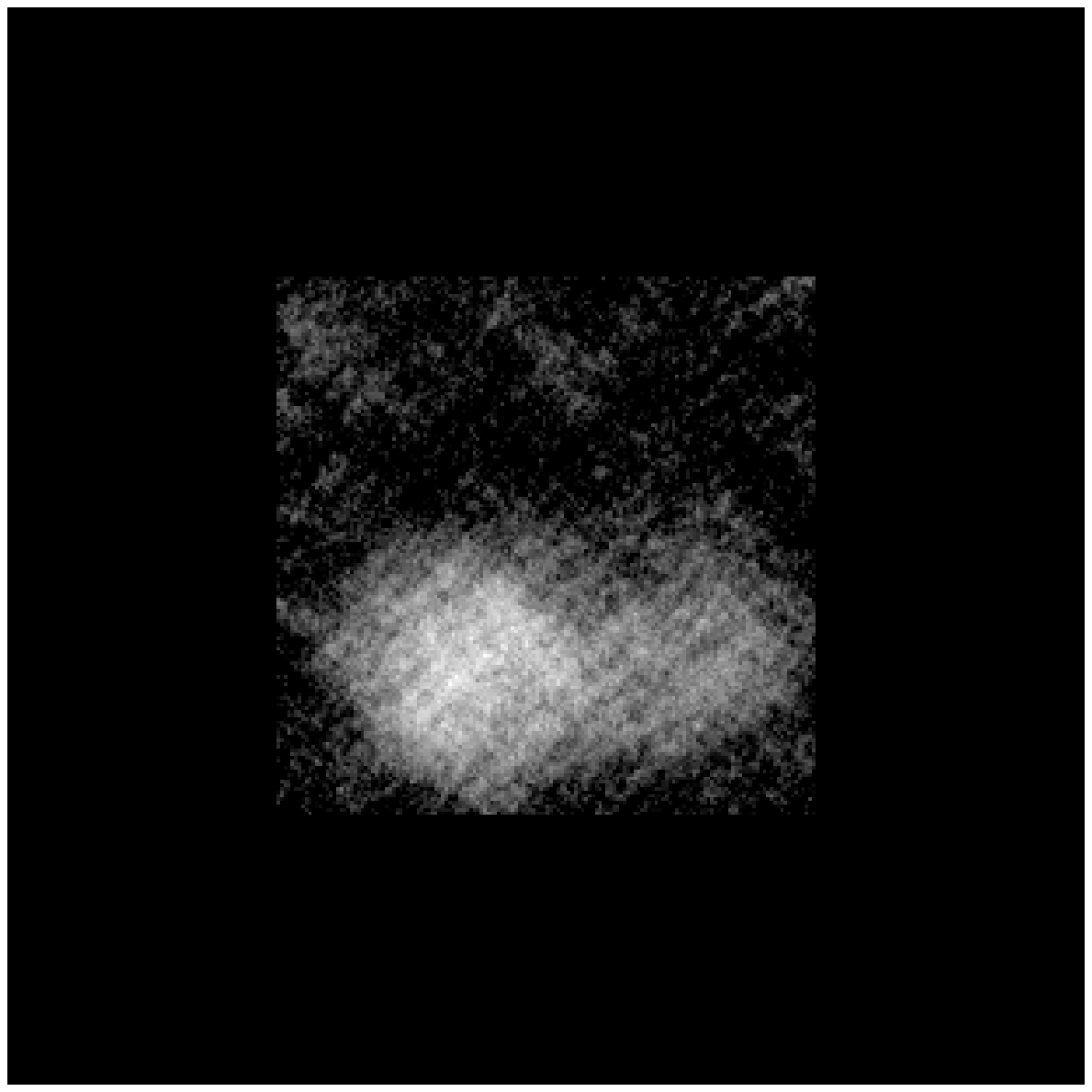}
  }
  \raisebox{-1.6pt}{
    \includegraphics[height=.282\figurewidth,bb = 470 238 536 596, clip]
    {colorbar.eps}}
  \caption{\label{fig:HIO}Reconstruction of the 2D electron density
    by the HIO method.}
\end{figure}

Figure~\ref{fig:error_F} shows the $\mathrm{Error}_{F}(\mu)$ for
diffraction images in Fig.~\ref{fig:lysozyme data}. The value of $\mu$
was varied from $1{\times}10^4$ to $1{\times}10^{-4}$. Note that
although the horizontal axis of Fig.~\ref{fig:error_F} increases from
left to right, the value of $\mu$ was decreased in the
experiments. The best $\mu$ was selected for each diffraction image by
$\mathrm{Error}_{F}(\mu)$ and the reconstructed electron densities are
shown in Figs.~\ref{fig:SPR}a, \ref{fig:SPR}b, and \ref{fig:SPR}c.

We also applied the HIO method for the data. The central square
($154\times{154}~\mathrm{pixels}$) was set as the support region. The
method has a problem with convergence, and we stopped the iteration
loop when the incremental difference of the estimated density becomes
small enough. Figure \ref{fig:HIO} shows the reconstructions obtained
by the HIO method for the diffraction images in Fig.~\ref{fig:lysozyme
  data}. The reconstructed electron densities are similar to
Fig.~\ref{fig:SPR}, but the SPR method obtained better
$\mathrm{Error}_{F}(\mu)$ for all of the diffraction images.

We note the computational time of the SPR method. The method was
implemented with C and run on a desktop machine with an Intel Xeon
X5570 (2.93~GHz 4~cores). The optimal electron density was computed
for all the values of $\mu$'s (33 different values of $\mu$) within 10
minutes.

\subsection{Missing centers}
\label{sec:missingcenter}

One of the problems of the phase retrieval from diffraction images is
missing centers. The intensities around the center cannot be measured
experimentally because the scattered beam is superimposed by the
direct beam around the center and a beam stop is used to block the
direct beam.

The SPR method can be applied to the case easily by modifying
eq.~\eqref{eq:proposed loss} as follows,
\begin{equation}
  \label{eq:proposed loss2}
  \ell_\mu(\V{f}|\V{N}) =\sum_{uv\in\alpha} \bigl(N_{uv} \ln
  |F_{uv}|^2-|F_{uv}|^2c_{uv}\bigr)
  -\mu\sum_{xy}w_{xy}f_{xy},
\end{equation}
where $\alpha$ is the active pixel set where the central pixels are
excluded. The error function is also modified as
\begin{equation}
  \label{eq:error_k2}
  \mathrm{Error}_{F}(\mu) =
  \frac{\sum_{uv\in\alpha}(|\hat{F}(\mu)_{uv}|{c_{uv}}^{1/2}-
    {N_{uv}}^{1/2})^2}{\sum_{u'v'\in\alpha}N_{u'v'}}.
\end{equation}

\begin{figure}[tbp]
  \centering 
  \includegraphics[width=.75\figurewidth]{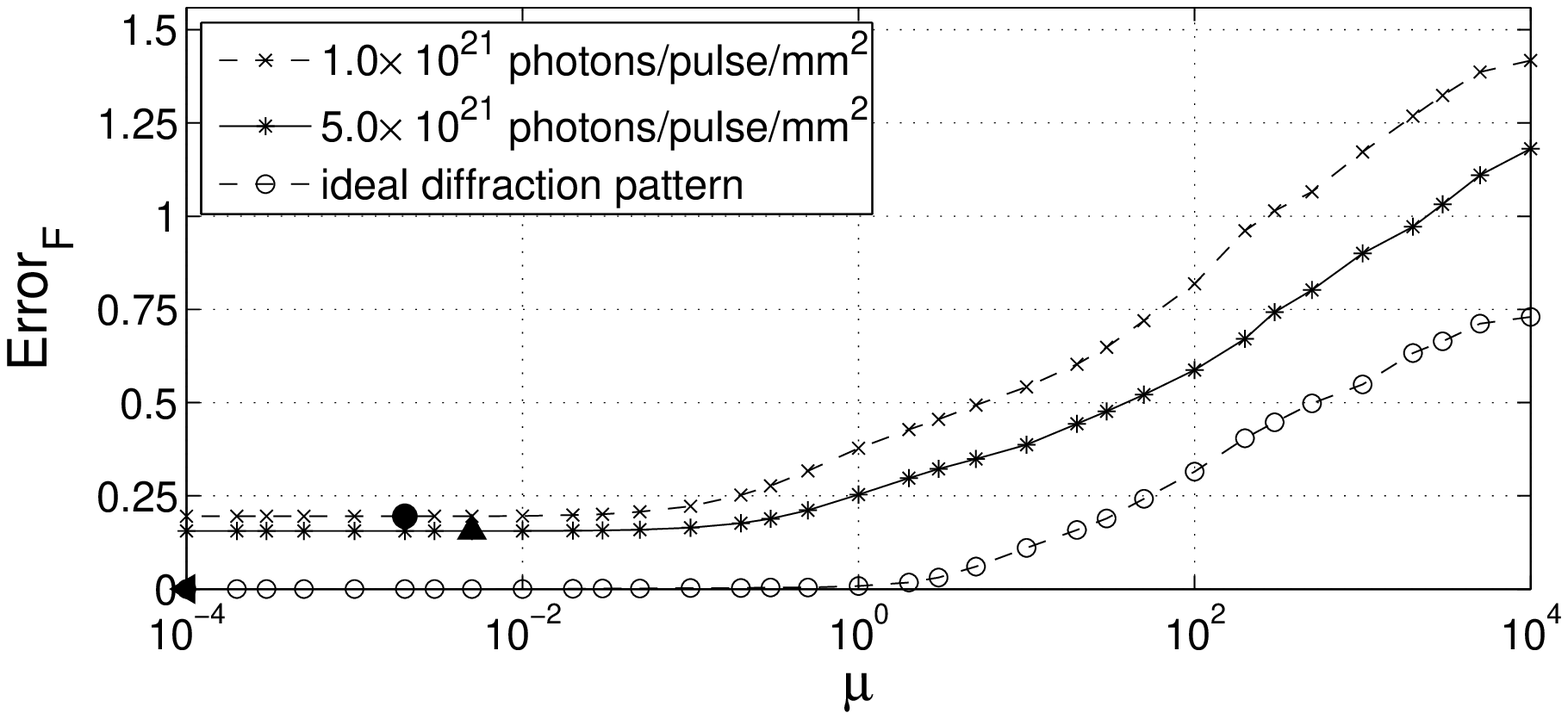}
  \caption{\label{fig:error_Fspot1}The values of
    $\mathrm{Error}_{F}(\mu)$ for simulated diffraction
    images in Fig.~\ref{fig:lysozyme data} without the central
    pixel. The black circle and triangles on the curves
    correspond to the reconstructed density images of
    Fig.~\ref{fig:SPRspot1}.}
\end{figure}

\begin{figure}[tb]
  \centering 
  \subfloat[Reconstruction from diffraction image
  Fig.2b without the central pixel 
    by the SPR method with $\mu=0.0001$.
  $\mathrm{Error}_{F}$ is 1.78$\times10^{-6}$.]{
      \includegraphics[width=.27\figurewidth]{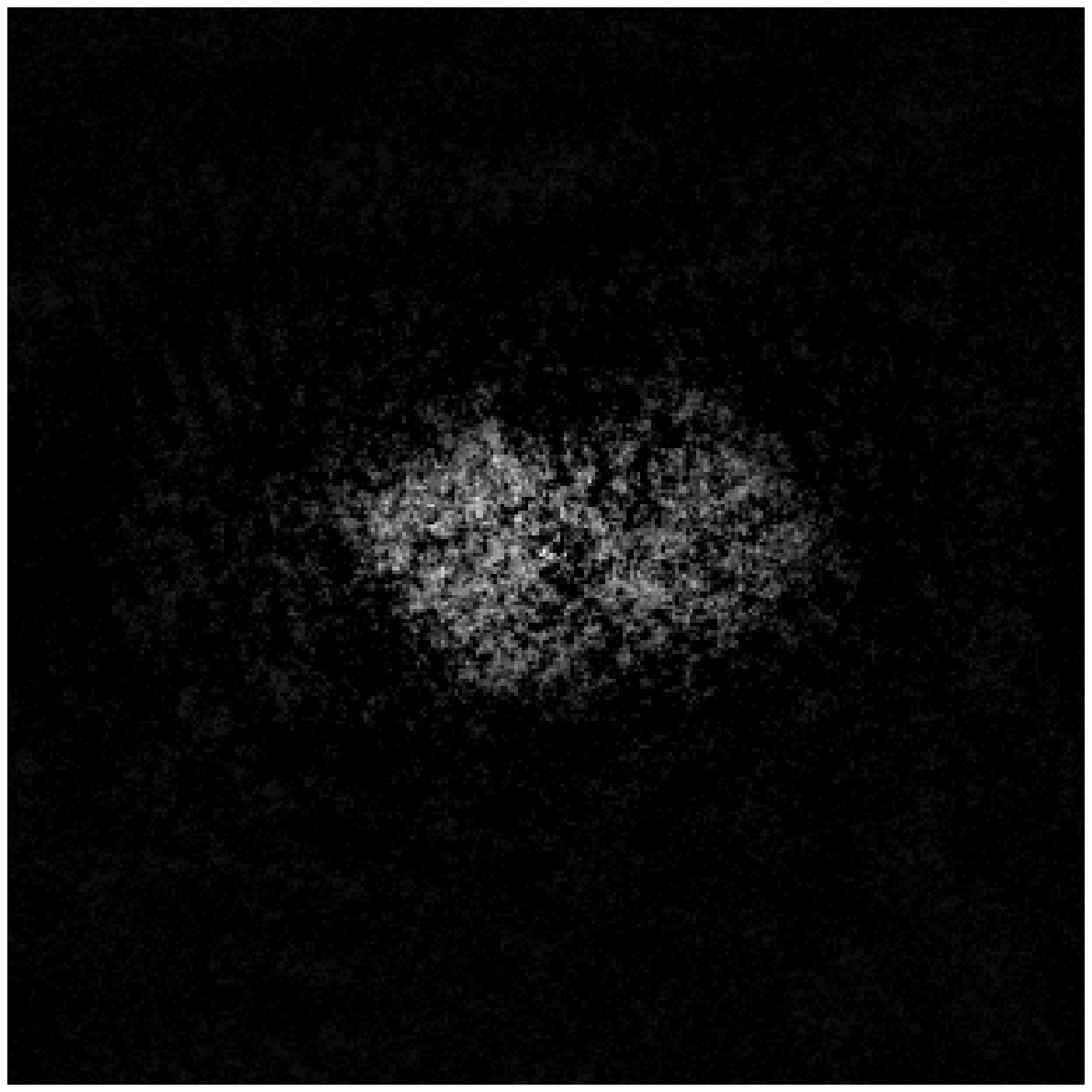}
    }\hfil
    \subfloat[Reconstruction from diffraction image
  Fig.2c without the central pixel
    by the SPR method with
  $\mu=0.002$. $\mathrm{Error}_{F}$ is 0.156.]{
      \includegraphics[width=.27\figurewidth]{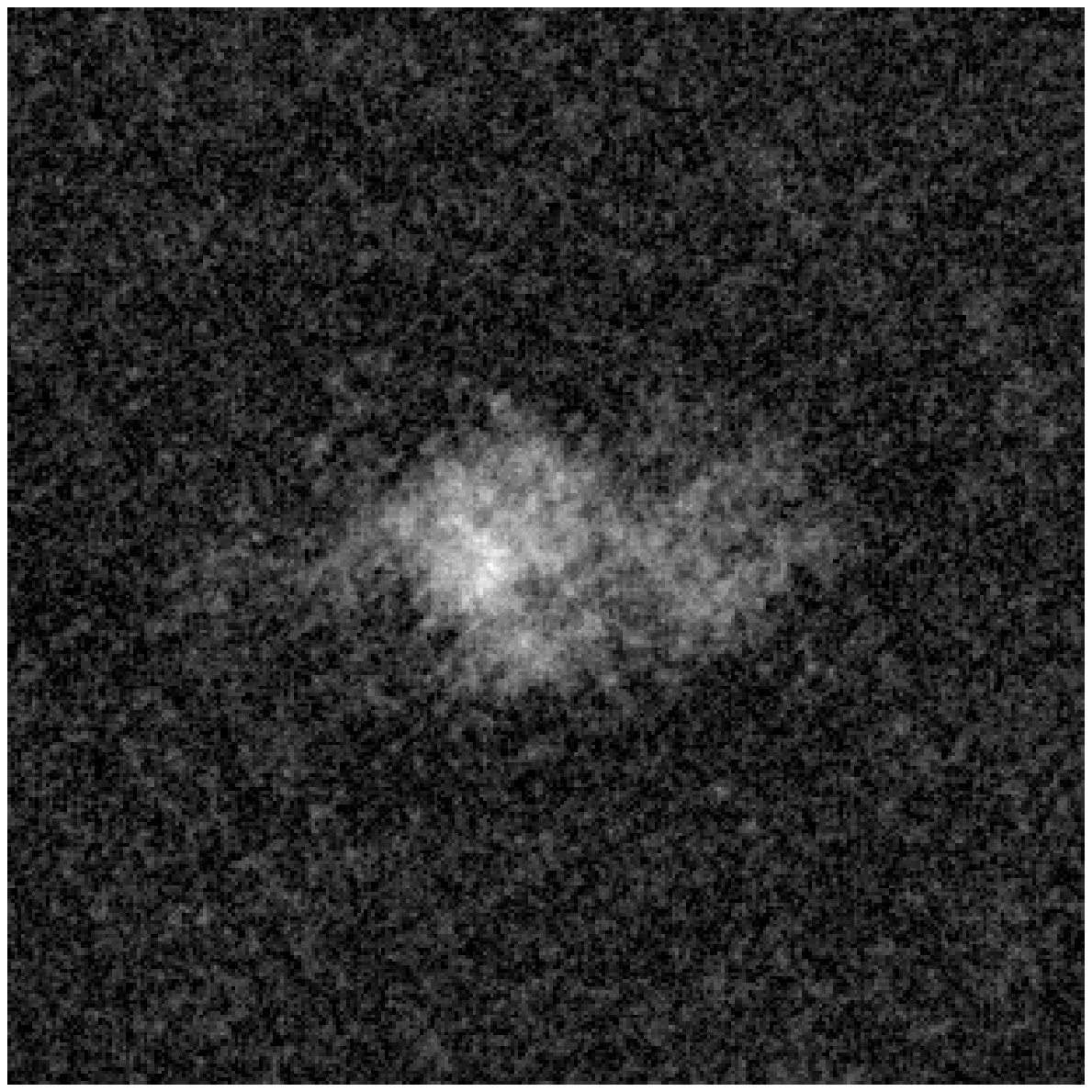}
    }\hfil
  \subfloat[Reconstruction from diffraction image
  Fig.2d without the central pixel
    by the SPR method with
  $\mu=0.005$. $\mathrm{Error}_{F}$ is 0.195.]{
    \includegraphics[width=.27\figurewidth]{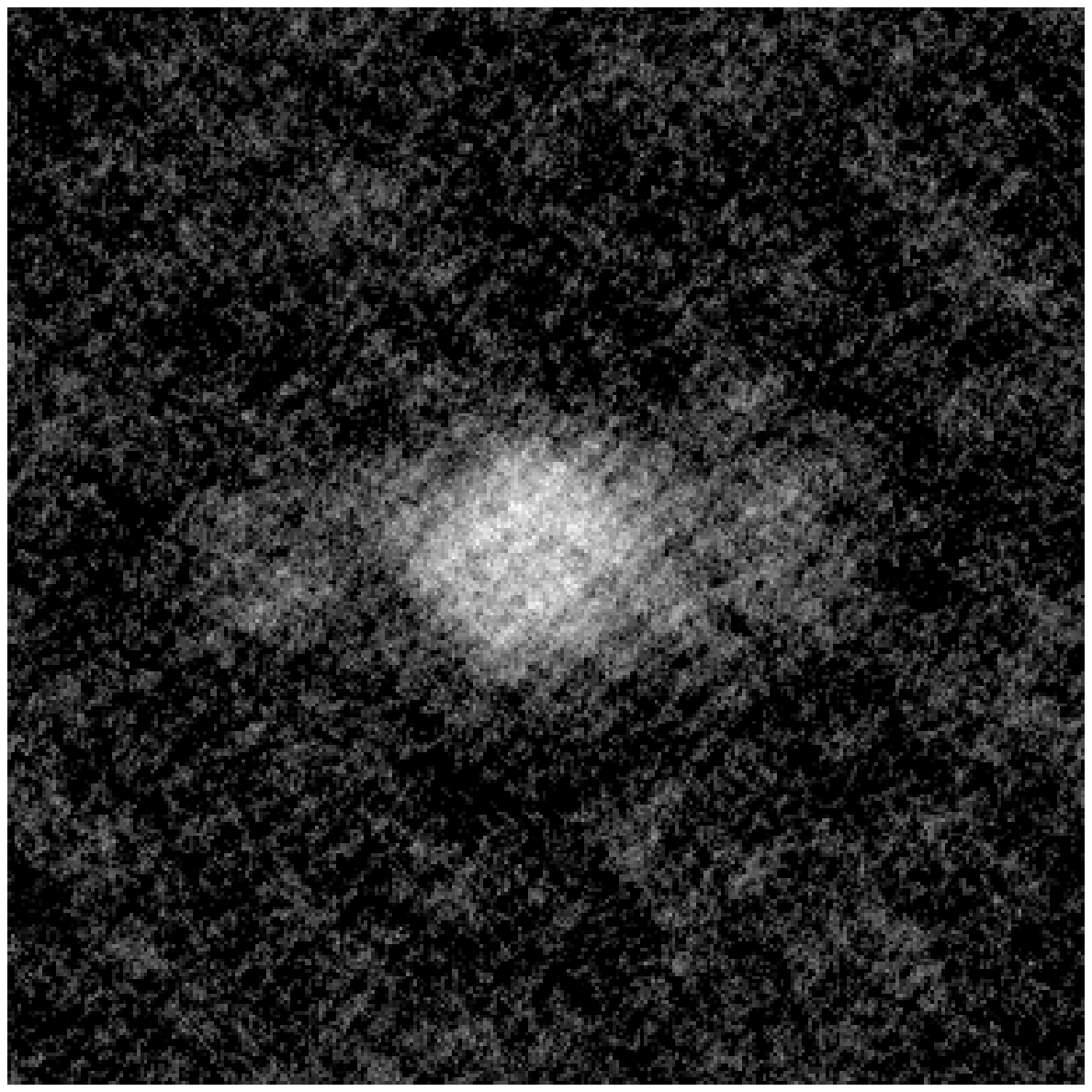}
  }
  \raisebox{-1.6pt}{
    \includegraphics[height=.282\figurewidth,bb = 470 238 536 596, clip]
    {colorbar.eps}}
    \caption{\label{fig:SPRspot1}Reconstruction of the 2D electron
      density by the SPR method.}
\end{figure}

\begin{figure}[htb]
  \centering 
  \subfloat[Reconstruction from diffraction image
  Fig.2b without the central pixel 
    by the HIO method.
  $\mathrm{Error}_{F}$ is 1.01$\times10^{-2}$.]{
      \includegraphics[width=.27\figurewidth]{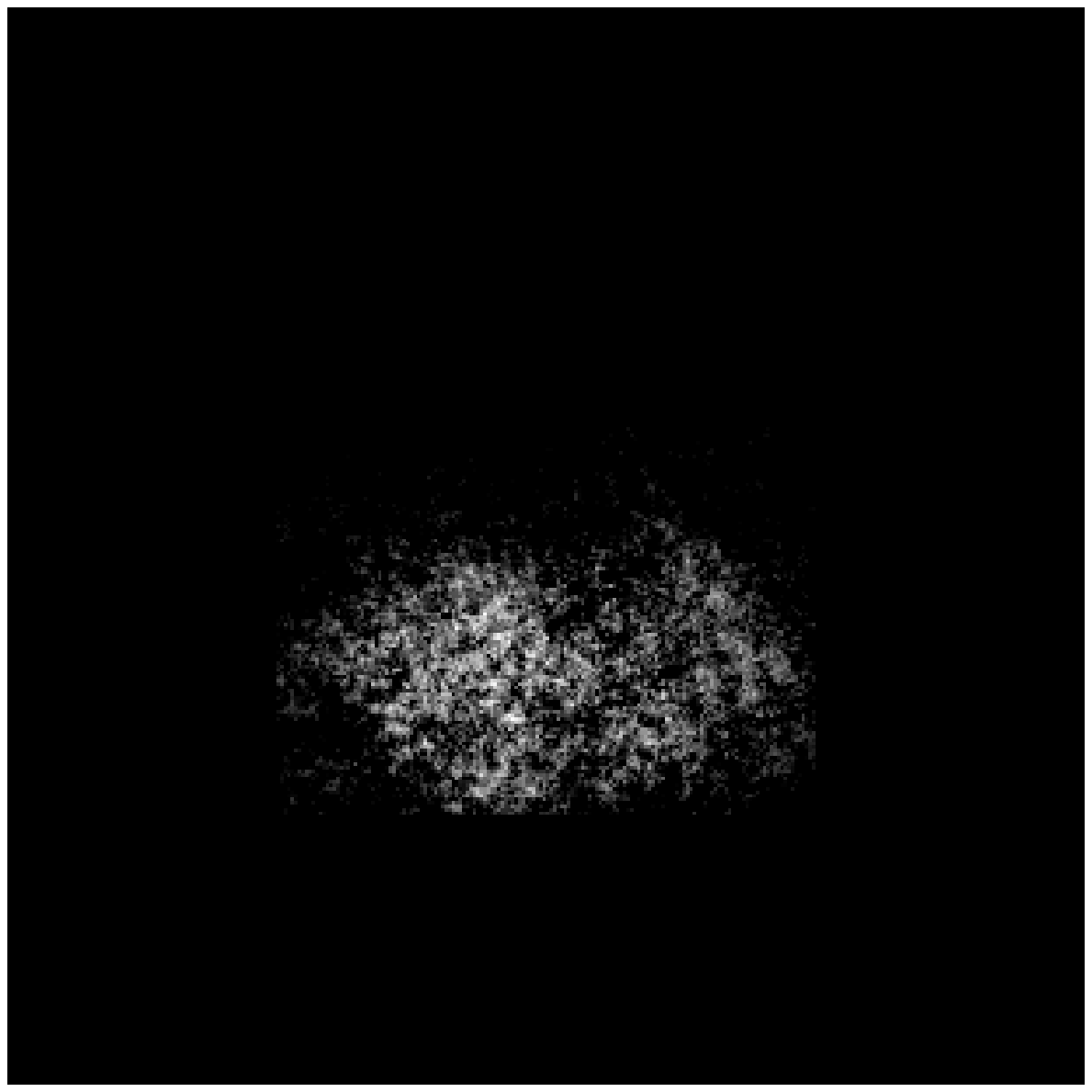}
    }\hfil
    \subfloat[Reconstruction from diffraction image
  Fig.2c without the central pixel
    by the HIO method. $\mathrm{Error}_{F}$ is 0.245.]{
      \includegraphics[width=.27\figurewidth]{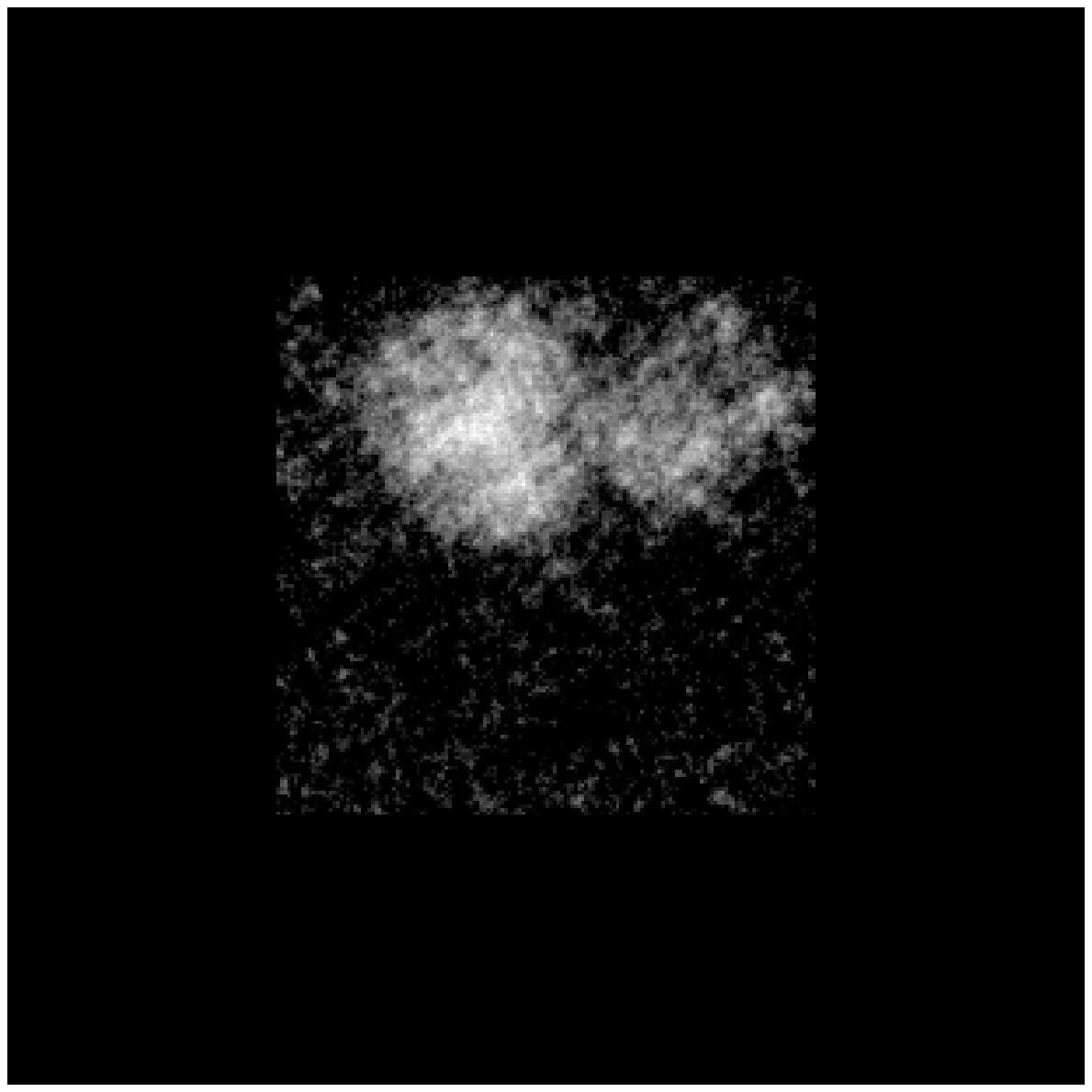}
    }\hfil
  \subfloat[Reconstruction from diffraction image
  Fig.2d without the central pixel
  by the HIO method. $\mathrm{Error}_{F}$ is 0.318.]{
    \includegraphics[width=.27\figurewidth]{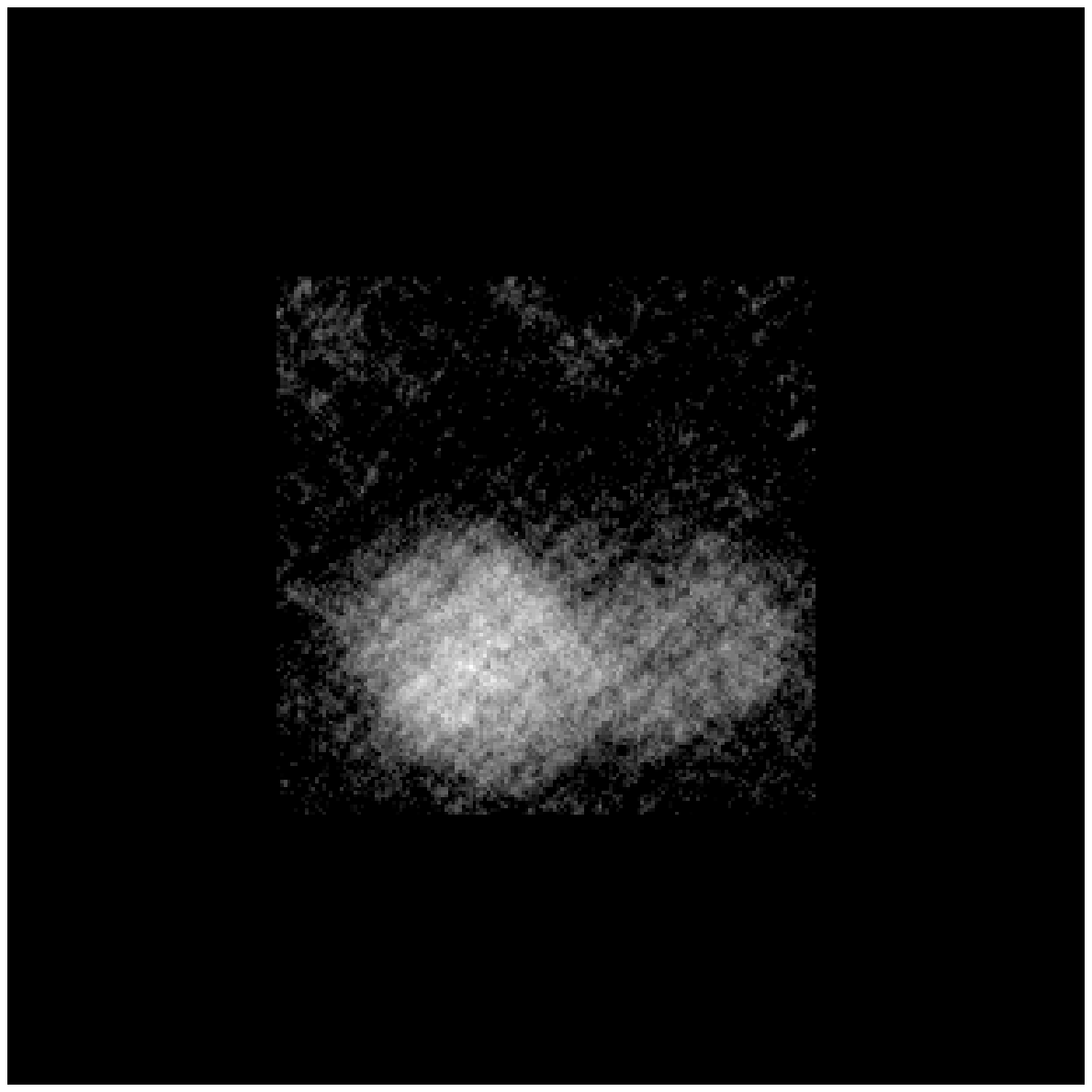}
  }
  \raisebox{-1.6pt}{
    \includegraphics[height=.282\figurewidth,bb = 470 238 536 596, clip]
    {colorbar.eps}}
  \caption{\label{fig:HIOspot1}Reconstruction of the 2D electron density
      by the HIO method.}
\end{figure}

Firstly, we masked the central pixel from each diffraction image in
Fig.~\ref{fig:lysozyme data} and applied the SPR method. By masking
the central pixel, Fig.~\ref{fig:lysozyme data}b lost around 5.7$\%$
of the photons, Fig.~\ref{fig:lysozyme data}c lost 294 out of 4630
photons and Fig.~\ref{fig:lysozyme data}d lost 56 out of
957 photons. Figure \ref{fig:error_Fspot1} shows the errors. Note that
the error values are better than those in Fig.~\ref{fig:error_F}, but
it does not mean the reconstructed densities in
Fig.~\ref{fig:SPRspot1} are better than those in Fig.~\ref{fig:SPR},
because the errors are computed only for the active pixel set. Figure
\ref{fig:HIOspot1} shows the results of the HIO method. Since the
central pixel corresponds to a scaling factor of the electron density
of the protein, the reconstructed images are quite similar to those
reconstructed from the complete diffraction images. We see that the
error values of the SPR method are better than those of the HIO
method.


\begin{figure}[p]
  \centering 
  \includegraphics[width=.75\figurewidth]{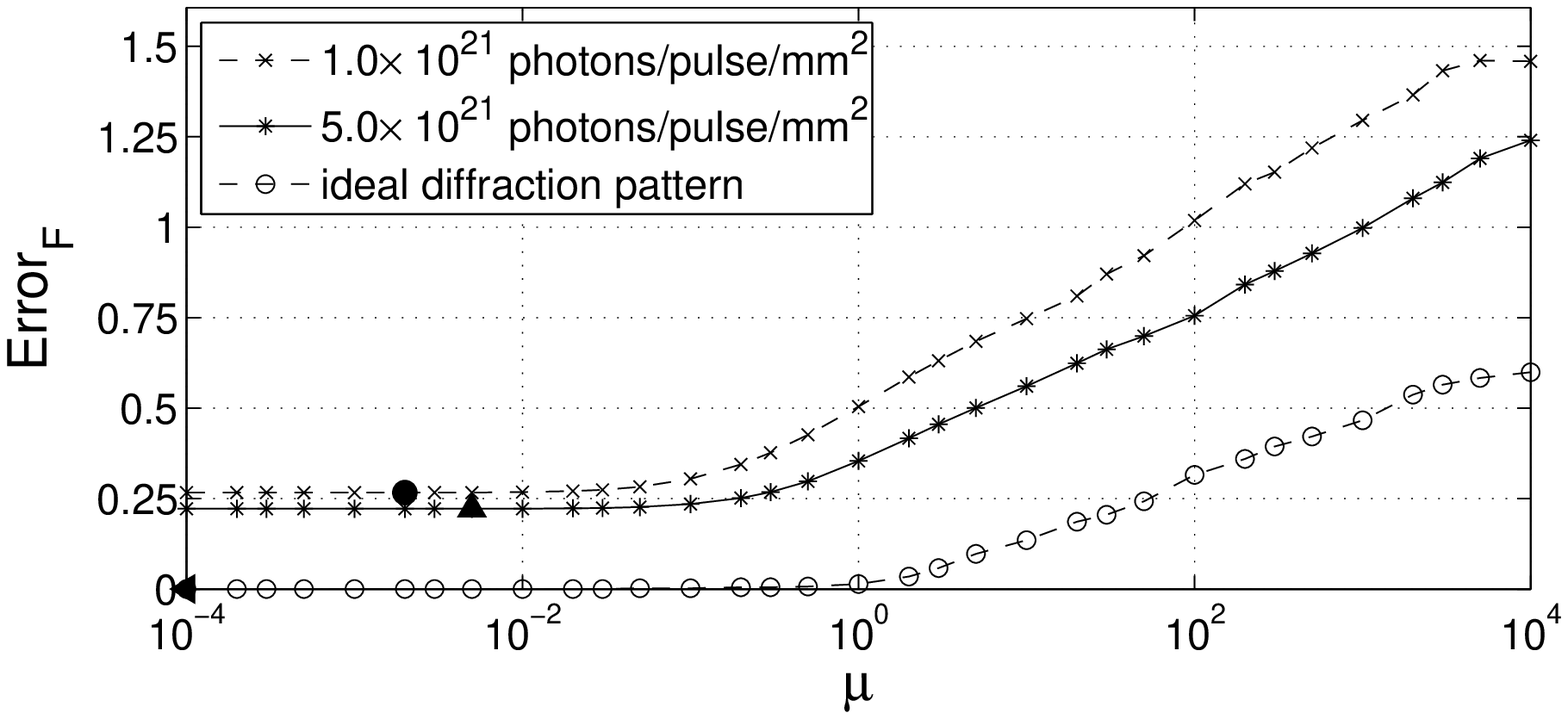}
  \caption{\label{fig:error_Fspot3}The values of
    $\mathrm{Error}_{F}(\mu)$ for simulated diffraction
    images in Fig.~\ref{fig:lysozyme data} without the central
    $3\times3$ pixels. The black circle and triangles on the curves
    correspond to the reconstructed density images of
    Figs.~\ref{fig:SPRspot3}.}
\end{figure}

\begin{figure}[tbp]
  \centering 
  \subfloat[Reconstruction from diffraction image
  Fig.2b without the central $3\times3$ pixels
  by the SPR method with $\mu=0.0001$.
  $\mathrm{Error}_{F}$ is 2.34$\times10^{-6}$.]{
      \includegraphics[width=.27\figurewidth]{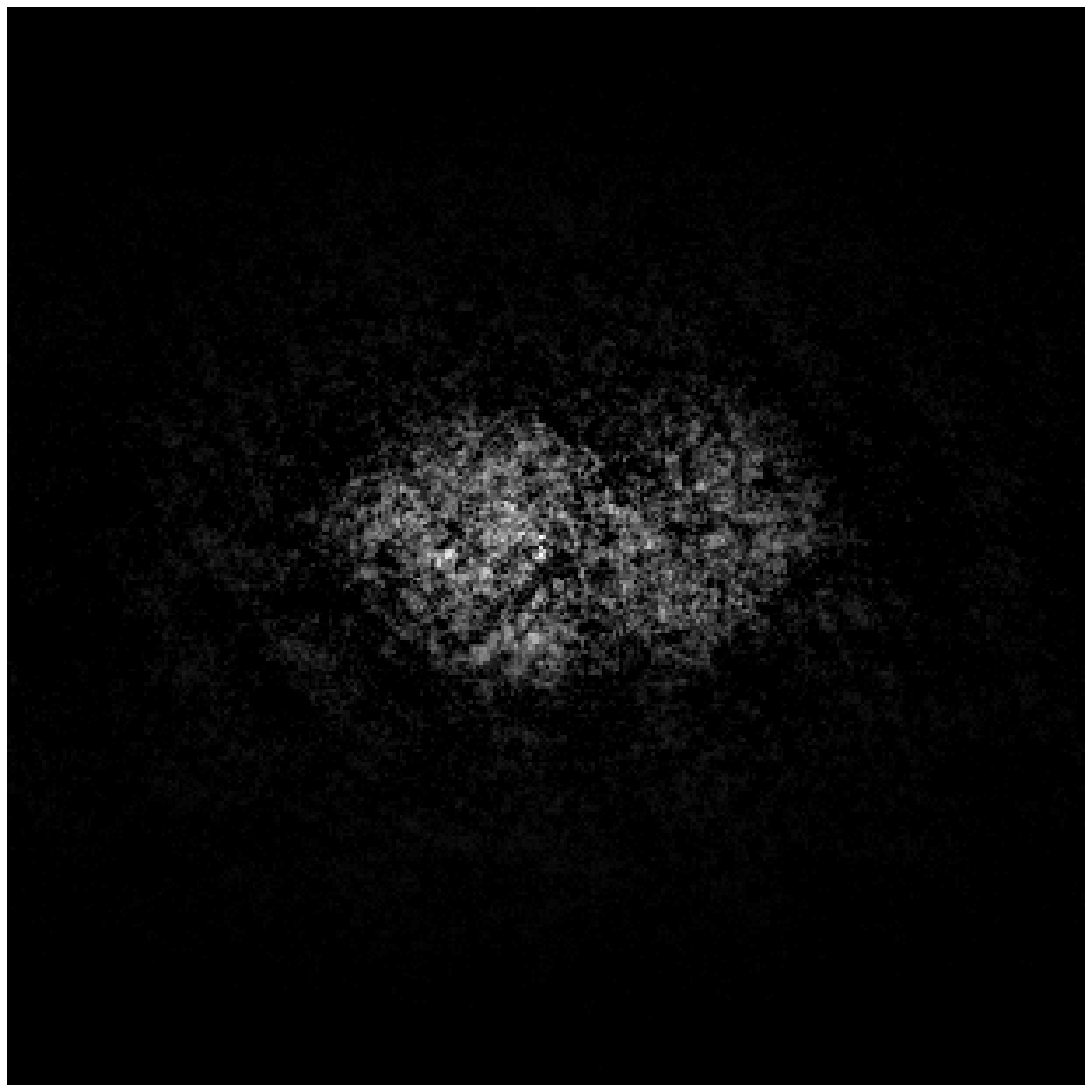}
    }\hfil
    \subfloat[Reconstruction from diffraction image
  Fig.2c without the central $3\times3$
  pixels by the SPR method with
  $\mu=0.002$. $\mathrm{Error}_{F}$ is 0.222.]{
      \includegraphics[width=.27\figurewidth]{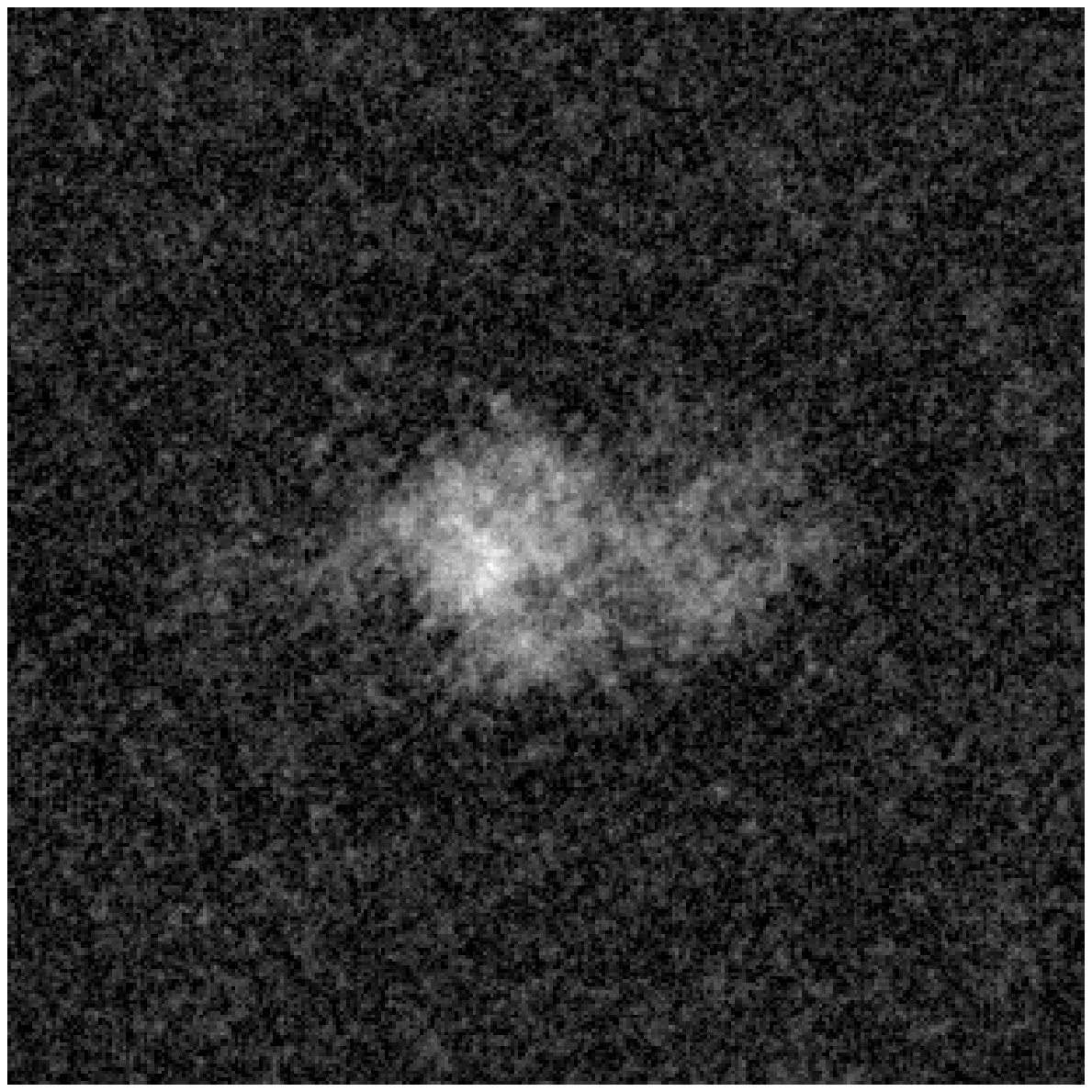}
    }\hfil
  \subfloat[Reconstruction from diffraction image
  Fig.2d without the central $3\times3$
  pixels by the SPR method with
  $\mu=0.005$. $\mathrm{Error}_{F}$ is 0.267.]{
    \includegraphics[width=.27\figurewidth]{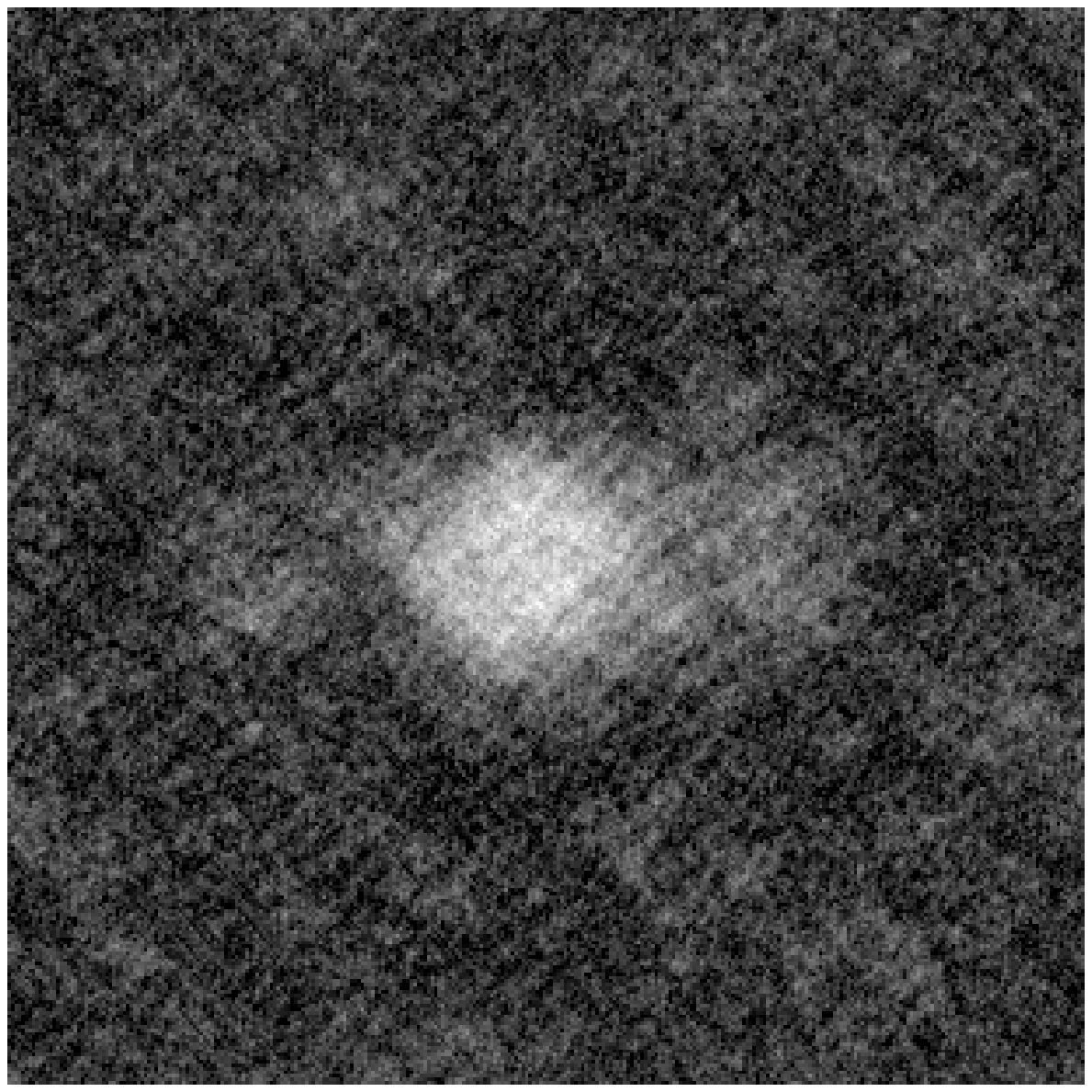}
  }
  \raisebox{-1.6pt}{
    \includegraphics[height=.282\figurewidth,bb = 470 238 536 596, clip]
    {colorbar.eps}}
    \caption{\label{fig:SPRspot3}Reconstruction of the 2D electron
      density by the SPR method.}
\end{figure}

\begin{figure}[tbp]
  \centering 
  \subfloat[Reconstruction from diffraction image Fig.2b without the
  central $3\times3$ pixels by the HIO method. $\mathrm{Error}_{F}$ is
  0.128.]{
      \includegraphics[width=.27\figurewidth]{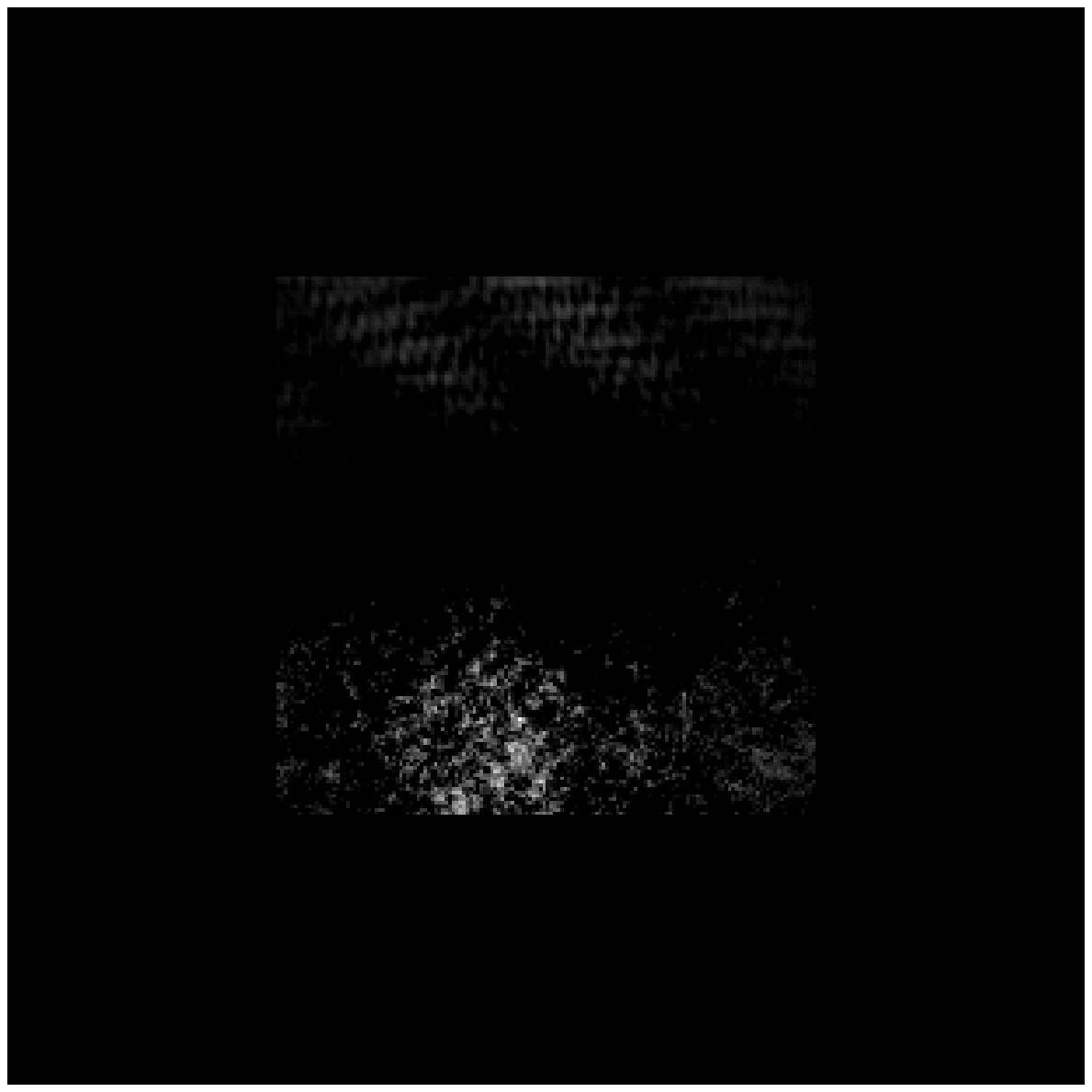}
    }\hfil
    \subfloat[Reconstruction from diffraction image Fig.2c without the
    central $3\times3$ pixels by the HIO method. $\mathrm{Error}_{F}$
    is 0.487.]{
      \includegraphics[width=.27\figurewidth]{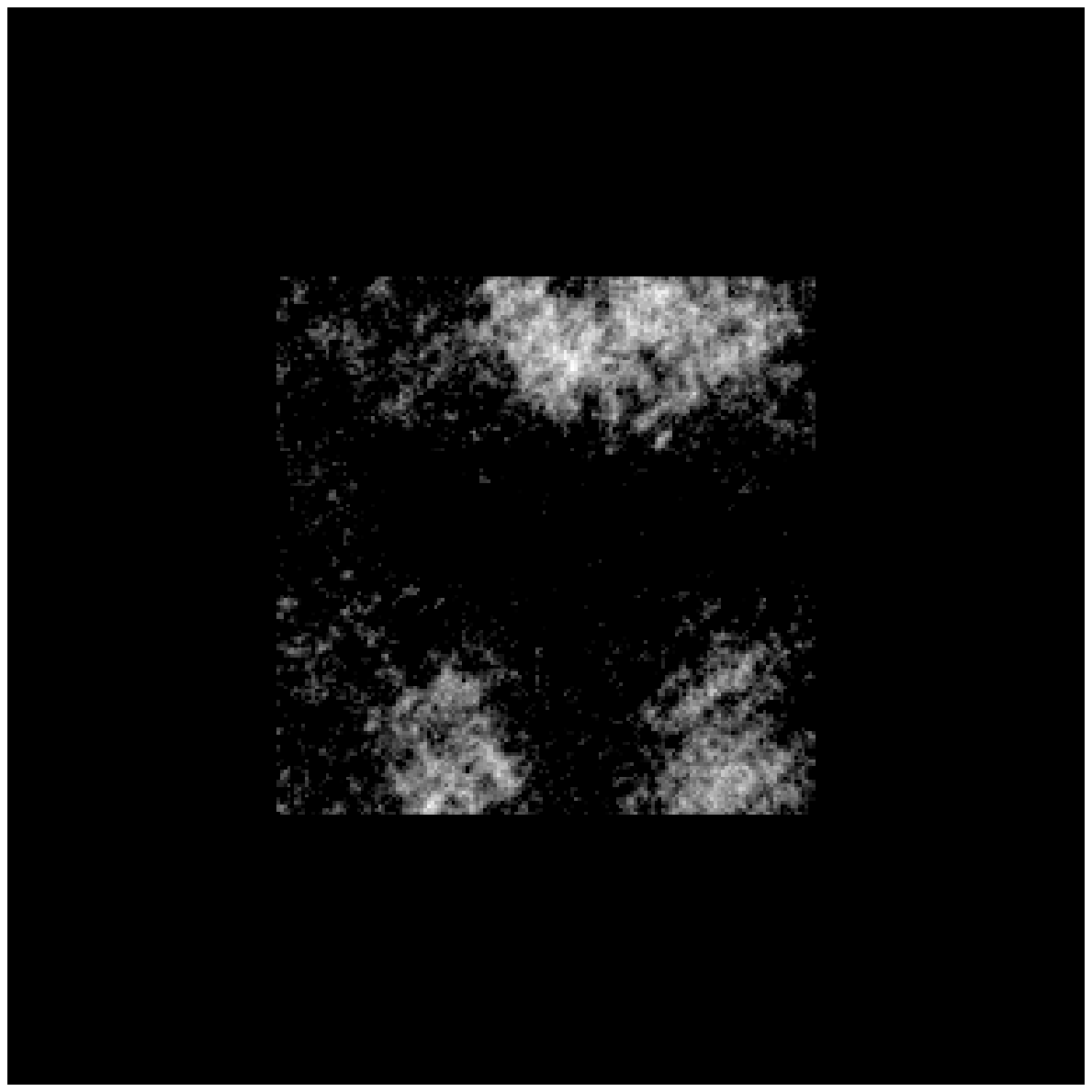}
    }\hfil
  \subfloat[Reconstruction from diffraction image Fig.2d without the
  central $3\times3$ pixels by the HIO method. $\mathrm{Error}_{F}$ is
  0.557.]{
    \includegraphics[width=.27\figurewidth]{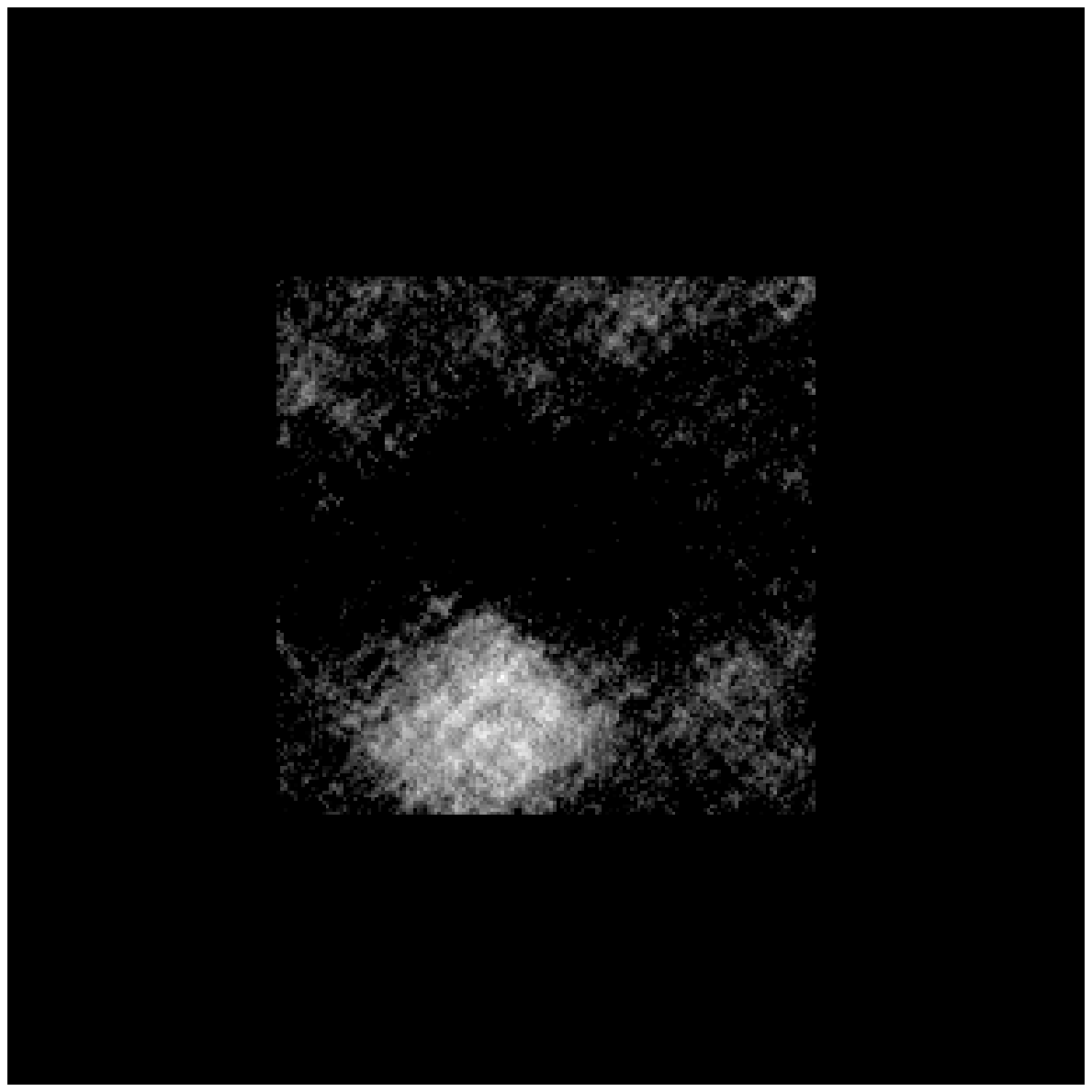}
  }
  \raisebox{-1.6pt}{
    \includegraphics[height=.282\figurewidth,bb = 470 238 536 596, clip]
    {colorbar.eps}}
  \caption{\label{fig:HIOspot3}Reconstruction of the 2D electron
    density by the HIO method.}
\end{figure}

Next, we further masked the central $3\times3$ pixels from each
diffraction image and applied the SPR method. Note that
Fig.~\ref{fig:lysozyme data}b lost around 35$\%$ of the photons,
Fig.~\ref{fig:lysozyme data}c lost 1595 out of 4630 photons and
Fig.~\ref{fig:lysozyme data}d lost 312 out of 957
photons. More than 30$\%$ of the photons are lost and the results are
expected to be degraded largely. Figure \ref{fig:error_Fspot3} shows
the errors. The reconstructed densities are shown in
Fig.~\ref{fig:SPRspot3}. Figure \ref{fig:HIOspot3} shows the results
of HIO method. We see that the error values of the SPR method are
better and the reconstructed results are more stable than the HIO
results.  It is worth noting that the SPR method worked well even if
the central $5\times 5$ pixels, a half area of the centrospeckle
pattern, were masked. In that condition, 54.4 $\%$ of the total photons
are lost.

\section{Discussion}
\label{sec:conclusion}

We have proposed a new phase retrieval method. The SPR method is based
on the MAP estimation of the Bayesian statistics, it will be effective
for single molecule diffraction images which will be obtained by
XFELs.

From a Bayesian viewpoint, the difference between other phase
retrieval methods including the HIO method and the SPR method is
summarized in the likelihood and the prior. Many methods use the
squared loss, which corresponds to a Gaussian likelihood, while the
present method employs a Poisson distribution. This is suitable for
CXDI data because the number of the scattered photons detected at each
pixel is a counting process. Similar idea has been found in a related
work~\cite{IrwanLane1998josaa}. The prior in HIO method corresponds to
a uniform distribution with a bounded support, while the proposed
method uses the exponential distribution. Similar prior has been
widely used in statistics~\cite{Tibshirani1996jrssB}. This combination
shows a new promising direction for phase retrieval.

The proposed method has been tested with simulated data. The electron
densities are reconstructed with a reasonable computational cost, and
errors of the results are smaller than the HIO method. We further
tested the method with the diffraction data where some central pixels
are masked. The proposed method is applied without any major
modification, and the results are quite stable even some pixels are
masked.

The SPR method reconstructs the density from an incomplete observation
utilizing the sparsity. This idea is similar to
CS~\cite{Donoho2006ieeeit}. In \cite{Marchesini2008arxiv}, CS is
applied to the phase retrieval problem, however, in our problem, the
number counts are very limited, and we needed to extend the model.

Finally, we list some of our future works to improve the SPR
method. One is the hyperparameter. The hyperparameter $\rho(\mu)_{xy}$
can be modified to reflect the knowledge of the true density. This may
result in a better estimate. Another issue is the algorithm. The
densities for the whole sequence of $\mu$ in Fig.~\ref{fig:error_F}
are computed with a simple line search algorithm but advanced
optimization methods may speed it up.

\section*{Acknowledgments}

We would like to thank Nobuhiro Go for fruitful discussion and Atsushi
Tokuhisa for providing the electron density data. We are also grateful
to Satoshi Ito for helpful discussion. This work was partly supported
by X-ray Free Electron Laser Utilization Research Project of the
Ministry of Education, Culture, Sports, Science and Technology of
Japan.


\begin{thebibliography}{10}

\bibitem{Sayre1980}
D.~Sayre.
\newblock Prospects for long-wavelength {X}-ray microscopy and diffraction.
\newblock In M.~Schlenker, M.~Fink, J.~P. Goedgebuer, C.~Malgrange, J.~Ch.
  Vi\'enot, and R.~H. Wade, editors, {\em Imaging Processes and Coherence in
  Physics}, volume 112 of {\em Springer Lecture Notes in Physics}, pages
  229--235. Springer, Berlin, 1980.

\bibitem{GaffneyChapman2007science}
K.~J. Gaffney and H.~N. Chapman.
\newblock Imaging atomic structure and dynamics with ultrafast x-ray
  scattering.
\newblock {\em Science}, 316:1444--1448, 2007.

\bibitem{Neutze_etal2000nature}
R.~Neutze, R.~Wouts, D.~van~der Spoel, E.~Weckert, and J.~Hajdu.
\newblock Potential for biomolecular imaging with femtosecond {X}-ray pulses).
\newblock {\em Nature}, 406:752--757, August 2000.

\bibitem{HuldtSzokeHajdu2003}
G.~Huldt, A.~Sz\H{o}ke, and J.~Hajdu.
\newblock Diffraction imaging of single particles and biomolecules.
\newblock {\em J. Str. Biol.}, 144:219--227, 2003.

\bibitem{Seibert_etal2011nature}
M.~Marvin Seibert~\emph{et al.}
\newblock Single mimivirus particles intercepted and imaged with an {X}-ray
  laser.
\newblock {\em Nature}, 470:78--81, February 2011.

\bibitem{BortelFaigel2007jsb}
G.~Bortel and G.~Faigel.
\newblock Classification of continuous diffraction patterns: A numerical study.
\newblock {\em J. Str. Biol.}, 158(1):10--18, April 2007.

\bibitem{BortelFaigelTegze2009jsb}
G.~Bortel, G.~Faigel, and M.~Tegze.
\newblock Classification and averaging of random orientation single
  macromolecular diffraction patterns at atomic resolution.
\newblock {\em J. Str. Biol.}, 166(3):226--233, 2009.

\bibitem{Fung_etal2009natphys}
R.~Fung, V.~Shneerson, D.~K. Saldin, and A.~Ourmazd.
\newblock Structure from fleeting illumination of faint spinning objects in
  flight.
\newblock {\em Nature Physics}, 5(1):64--67, January 2009.

\bibitem{LohVeit2009pre}
N.-T.~D. Loh and V. Elser.
\newblock Reconstructon algorithm for single-particle diffraction imaging
  experiments.
\newblock {\em Physical Review E}, 80:026705, 2009.

\bibitem{Saldin_etal2009iop}
D.~K. Saldin, V.~L. Shneerson, R.~Fung, and A.~Ourmazd.
\newblock Structure of isolated biomolecules obtained from ultrashort x-ray
  pulses: exploiting the symmetry of random orientations.
\newblock {\em J. of Physics: Condensed Matter}, 21(13):134014, 2009.

\bibitem{Frank2006oup}
J.~Frank.
\newblock {\em Three-Dimensional Electron Microscopy Of Macromolecular
  Assemblies: Visualization Of Biological Molecules In Their Native State}.
\newblock Oxford University Press, Oxford, 2006.

\bibitem{Young_etal2010nature}
L.~{Young \emph{et al.}}
\newblock Femtosecond electronic response of atoms to ultra-intense {X}-rays.
\newblock {\em Nature}, 466:56--61, 2010.

\bibitem{Tibshirani1996jrssB}
R.~Tibshirani.
\newblock Regression shrinkage and selection via the lasso.
\newblock {\em J. R. Stat. Soc., Ser. B}, 58(1):267--288, 1996.

\bibitem{ZhaoYu2006jmlr}
P.~Zhao and B.~Yu.
\newblock On model selection consistency of lasso.
\newblock {\em J. Machine Learning Research}, 7:2541--2563, 2006.

\bibitem{Marchesini2008arxiv}
S.~Marchesini.
\newblock Ab initio compressive phase retrieval.
\newblock {\em ArXiv Physics Optics e-prints}, 2008.

\bibitem{IrwanLane1998josaa}
R.~Irwan and R.~G. Lane.
\newblock Phase retrieval with prior information.
\newblock {\em J. Opt. Soc. Am. A}, 15(9):2302--2311, 1998.

\bibitem{BaskaranMillane1999ieeeip}
S.~Baskaran and R.~P. Millane.
\newblock Bayesian image reconstruction from partial image and aliased spectral
  intensity data.
\newblock {\em IEEE trans. Image Proc.}, 8(10):1420--1434, 1999.

\bibitem{Fienup1978optlett}
J.~R. Fienup.
\newblock Reconstruction of an object from the modulus of its {F}ourier
  transform.
\newblock {\em Optics Letters}, 3(1):27--29, 1978.

\bibitem{Fienup1982appopt}
J.~R. Fienup.
\newblock Phase retrieval algorithms: a comparison.
\newblock {\em Applied Optics}, 21(15):2758--2769, August 1982.

\bibitem{MiaoSayreChapman1998josaa}
J.~Miao, D.~Sayre, and H.~N. Chapman.
\newblock Phase retrieval from the magnitude of the {F}ourier transforms of
  nonperiodic objects.
\newblock {\em J. Opt. Soc. Am. A}, 15(6):1662--1669, 1998.

\bibitem{Donoho2006ieeeit}
D.~Donoho.
\newblock Compressed sensing.
\newblock {\em IEEE transaction on Information Theory}, 52(4):1289--1306, April
  2006.

\bibitem{ZouHastieTibshirani2007as}
H.~Zou, T.~Hastie, and R.~Tibshirani.
\newblock On the ``degrees of freedom'' of the lasso.
\newblock {\em Annals of Statistics}, 35(5):2173--2192, 2007.

\end{thebibliography}

\end{document}